# New three-dimensional strain-rate potentials for isotropic porous metals: role of the plastic flow of the matrix

Benoit Revil-Baudard and Oana Cazacu[*]

Department of Mechanical and Aerospace Engineering, University of Florida, REEF, 1350 N. Poquito Rd., Shalimar, FL 32579, USA.

**ABSTRACT**

At present, modeling of the plastic response of porous solids is done using stress-based plastic potentials. To gain understanding of the combined effects of all invariants for general three-dimensional loadings, a strain-rate based approach appears more appropriate. In this paper, for the first time strain rate-based potentials for porous solids with Tresca and von Mises, matrices are obtained. The dilatational response is investigated for general 3-D conditions for both compressive and tensile states using rigorous upscaling methods. It is demonstrated that the presence of voids induces dependence on all invariants, the noteworthy result being the key role played by the plastic flow of the matrix on the dilatational response. If the matrix obeys the von Mises criterion, the shape of the cross-sections of the porous solid with the octahedral plane deviates slightly from a circle, and changes very little as the absolute value of the mean strain rate increases. However, if the matrix behavior is described by Tresca's criterion, the shape of the cross-sections evolves from a regular hexagon to a smooth triangle with rounded corners. Furthermore, it is revealed that the couplings between invariants are very specific and depend strongly on the particularities of the plastic flow of the matrix.

[*] Corresponding author: Tel: +1 850 833 9350; fax: +1 850 833 9366.
E-mail address: cazacu@reef.ufl.edu

# 1. INTRODUCTION

Using the plastic work equivalence principle, Ziegler (1977) has shown that a strain-rate potential can be associated to any convex stress-potential. Thus, to describe plastic flow a strain-rate potential (SRP) can be used instead of a stress-based potential. Such an approach is generally adopted in crystal plasticity because it is much easier to calculate numerically a crystallographic strain-rate potential than to compute a crystallographic yield surface (e.g. see Van Bael and Van Houtte, 2004).

Analytic expressions of macroscopic strain-rate potentials associated to stress-based potentials are known only for classical isotropic yield criteria such as Mises, Tresca, or Drucker-Prager (see Salencon, 1983), the orthotropic Hill (1948) criterion (see Hill, 1987), the orthotropic criterion of Cazacu et al. (2006) (see Cazacu et al. 2010; Yoon et al., 2011). If the stress-based potentials are non-quadratic obtaining an analytic expression for the exact dual is very challenging, if not impossible. Barlat and co-workers have proposed several analytic non-quadratic anisotropic strain rate potentials (e.g. Barlat et al., 2003; Kim et al., 2007). Although none of these strain rate potentials are strictly dual (conjugate) of the respective stress potentials, it was shown that these strain rate formulations lead to a description of the plastic anisotropy of FCC metals of comparable accuracy to that obtained using the stress potentials. It is generally agreed that strain-rate potentials are more suitable for process design involving the solution of inverse problems (see Barlat et al., 1993; Chung et al., 1997).

It is to be noted that all the strain-rate based potentials mentioned above apply only to fully-dense materials for which the plastic flow is incompressible. However, all engineering materials contain defects (cracks, voids). Based on micromechanical considerations, Rice and Tracey (1969) and Gurson (1977) have demonstrated that the presence of voids induces a dependence of the mechanical response on the mean stress and as such the plastic flow is accompanied by volume changes. Stress-based potentials have been developed to capture the characteristics of the plastic flow of porous metallic materials. Most of these models are based on the hypothesis that the matrix (void-free material) obeys von Mises yield criterion. This is the case, for example, of the classical Gurson (1977) model and its various extensions proposed by Tvergaard (1981), Gologanu (1997), Tvergaard and Nielsen (2010), Lecarme et al. (2011), among others. It is to be noted that in all these Gurson-type models as well as in recent phenomenological models (e.g. Stoughton and Yoon, 2011; Yoon et al. 2013, etc.) the effects of the mean stress and shear stresses are decoupled. However, finite-element (FE) cell model calculations as well as very recent full-field calculations of the yield surface of voided polycrystals deforming by slip have shown a very specific dependence of yielding with the signs of the mean stress and the third-invariant of the stress deviator, $J_3^\Sigma$. Specifically, for tensile loadings the response corresponding to $J_3^\Sigma \geq 0$ is softer than that corresponding to $J_3^\Sigma \leq 0$ while for compressive loadings, the reverse occurs (e.g. Richelsen and Tvergaard, 1994; Cazacu and Stewart, 2009; Lebensohn and Cazacu, 2012). For axisymmetric loadings, using micromechanical considerations Cazacu et al. (2013; 2014) developed analytical yield criteria for porous materials with von Mises and Tresca matrix, respectively. These stress-based potentials account for the combined effects of the sign of the mean stress and of the third-invariant of the stress deviator on the dilatational response. Most importantly, it was explained the role of the third-invariant on void growth and void collapse (see Alves et al., 2014). However, to gain understanding of the combined effects of all invariants for general three-dimensional loadings, a strain-rate based approach appears more appropriate.

The need for models that can explain and better predict damage accumulation and its influence on the plastic response of engineering materials was evidenced by recent experimental studies where the mechanical response for full three-dimensional loadings of both polycrystalline metallic materials (e.g. Barsoum and Faleskog, 2007; Maire et al., 2011; Khan and Liu, 2012, etc.) and microcellular metallic materials (or metallic ''foams'' (e.g. Combaz et al., 2011)). In particular, Combaz et al. (2011) reported data on aluminum foams for multi-axial loading paths corresponding to different values of $J_3^\Sigma$ for both tensile and compressive loadings, showing that the yield locus is centro-symmetric, its shape in the octahedral plane being triangular.

In this paper, for the first time three-dimensional (3-D) strain-rate potentials for porous solids with von Mises and Tresca matrix are obtained and their properties investigated for both compressive and tensile loadings. The structure of the paper is as follows. We begin with a brief presentation of the modeling framework (Section 2). The strain-rate potentials for porous von Mises and Tresca solids are given in Section 3. To fully assess the properties of the respective potentials, the shape of the cross-sections with the octahedral plane are analyzed. For a porous solid with von Mises matrix, the noteworthy result is that the shape of the cross-section changes little with the mean strain rate and the most pronounced influence of the third-invariant occurs for axisymmetric states

However, if the plastic flow in the matrix is governed by Tresca's criterion, there is a very strong coupling between all invariants. New features of the dilatational response are revealed. Irrespective of the level of porosity, the shape of the cross-section in the octahedral plane changes from a regular hexagon to a rounded triangle, the surface closing on the hydrostatic axis. Furthermore, it is shown that the level of porosity in the material strongly influences the couplings between the invariants. In Section 4, we revisit some aspects of Gurson's treatment. More specifically, the implications of the approximations made by Gurson (1975; 1977) and the properties of the strain-rate potential, which is the exact conjugate of its classical stress-based potential (Gurson, 1977)) are discussed in relation to the exact strain-rate potentials for porous solids with Tresca and von Mises matrix presented in this study (Section 4). The main findings of this paper are summarized in Section 5.

Regarding notations, vector and tensors are denoted by boldface characters. If **A** and **B** are second-order tensors, the contracted tensor product between such tensors is defined as: $\mathbf{A}:\mathbf{B} = A_{ij}B_{ij}$ $i, j = 1\ldots 3$.

## 2. Modeling framework

### 2.1. Strain-rate potentials for fully-dense metallic materials

Generally, the onset of plastic flow is described by specifying a convex yield function, $\varphi(\boldsymbol{\sigma})$, in the stress space and the associated flow rule

$$\mathbf{d} = \dot{\lambda}\frac{\partial \varphi}{\partial \boldsymbol{\sigma}}, \qquad (1)$$

where $\boldsymbol{\sigma}$ is the Cauchy stress tensor, $\mathbf{d}$ denotes the plastic strain rate tensor and $\dot{\lambda} \geq 0$ stands for the plastic multiplier. The yield surface is defined as: $\varphi(\boldsymbol{\sigma}) = \sigma_T$, where $\sigma_T$ is the uniaxial yield in tension. The dual potential of the stress potential $\varphi(\boldsymbol{\sigma})$ is defined (see Ziegler, 1977; Hill, 1987) as:

$$\psi(\mathbf{d}) = \dot{\lambda}, \qquad (2)$$

and

$$\boldsymbol{\sigma} = \sigma_T \frac{\partial \psi}{\partial \mathbf{d}}. \qquad (3)$$

The yield function $\varphi(\boldsymbol{\sigma})$ is generally taken homogeneous of degree one with respect to positive multipliers, so the plastic dissipation is:

$$\pi(\mathbf{d}) = \sup_{\boldsymbol{\sigma} \in \mathcal{C}}\left(\sigma_{ij} d_{ij}\right) = \dot{\lambda}\sigma_T, \qquad i, j = 1\ldots 3, \qquad (4)$$

where $\mathcal{C}$ is the convex domain delimited by the yield surface. Thus, the functions $\psi(\mathbf{d})$ and $\varphi(\boldsymbol{\sigma})$ are dual potentials. For example, in the case of the von Mises potential, i.e.

$$\varphi_{\text{Mises}}(\boldsymbol{\sigma}) = \sqrt{(3/2)\boldsymbol{\sigma}':\boldsymbol{\sigma}'},$$

the associated strain-rate potential is: $\psi_{\text{Mises}}(\mathbf{d}) = \sqrt{(2/3)\mathbf{d}:\mathbf{d}} = \dot{\bar{\varepsilon}}$, where $\dot{\bar{\varepsilon}}$ denotes the von Mises equivalent strain rate and $\boldsymbol{\sigma}'$ the stress deviator, and the plastic dissipation is

$$\pi_{\text{Mises}}(\mathbf{d}) = \sigma_T \sqrt{(2/3)\mathbf{d}:\mathbf{d}}. \qquad (5)$$

In the case of Tresca's stress potential, i.e. $\varphi(\boldsymbol{\sigma}) = \max(|\sigma_1 - \sigma_2|, |\sigma_2 - \sigma_3|, |\sigma_1 - \sigma_3|)$, with $\sigma_1$, $\sigma_2$, and $\sigma_3$, being the principal values of $\boldsymbol{\sigma}$, the associated strain-rate potential is: $\psi_{\text{Tresca}}(\mathbf{d}) = (|d_1| + |d_2| + |d_3|)/2$, with $d_1$, $d_2$ and $d_3$ being the principal values of $\mathbf{d}$, and the plastic dissipation is:

$$\pi_{\text{Tresca}}(\mathbf{d}) = \frac{\sigma_T}{2}(|d_1| + |d_2| + |d_3|) \tag{6}$$

For an isotropic material (which means that $\varphi(\boldsymbol{\sigma})$ of Eq.(1) is an isotropic function), the principal directions of $\mathbf{d}$ and $\boldsymbol{\sigma}$ coincide. As an example, in Figure 1(a) are shown the projections of the strain-rate potentials $\psi_{\text{Tresca}}$ and $\psi_{\text{Mises}}$, respectively in any deviatoric plane (i.e. a plane with normal at equal angles to the principal directions of the plastic strain rate tensor $\mathbf{d}$). Figure 1(b) depicts the cross-sections of the classical Tresca and von Mises yield surfaces (normalized by $\sigma_T$), respectively. Since both Tresca and von Mises criteria are pressure-insensitive, irrespective of the value of the mean stress, $\sigma_m = (\sigma_1 + \sigma_2 + \sigma_3)/3$, the projection in the octahedral plane of the Mises surface (cylinder) is always a circle, while the projection of Tresca's yield surface (prism) is a regular hexagon. Since the von Mises yield surface is an upper bound for Tresca's, the Mises circle circumscribes the Tresca's hexagon (see Fig 1(b)). Each SRP being the exact conjugate of the respective stress potential, the Tresca SRP is an upper-bound of the von Mises SRP (see Fig 1(a)).

## 2.2. Kinematic homogenization framework for development of plastic potentials for porous metallic materials

The kinematic homogenization approach of Hill-Mandel (Hill, 1967; Mandel, 1972) offers a rigorous framework for the development of plastic potentials for porous solids. If the matrix (void-free material) is rigid-plastic, it has been shown (see Talbot and Willis, 1985) that there exists a strain-rate potential $\Pi = \Pi(\mathbf{D}, f)$ such that the stress at any point in the porous solid is given by:

$$\boldsymbol{\Sigma} = \frac{\partial \Pi(\mathbf{D}, f)}{\partial \mathbf{D}} \text{ with } \Pi(\mathbf{D}, f) = \inf_{\mathbf{d} \in K(\mathbf{D})} \langle \pi(\mathbf{d}) \rangle_\Omega, \tag{7}$$

where $\Omega$ is a representative volume element composed of the matrix and a traction-free void, while $\langle \rangle$ denotes the average value over $\Omega$; $f$ is the porosity (ratio between the volume of the void and the volume of $\Omega$); $\pi(\mathbf{d})$ is the matrix's plastic dissipation with $\mathbf{d}$ being the local plastic strain rate tensor. In Eq. (7), inf stands for infimum, the minimization being done over $K(\mathbf{D})$, which is the set of incompressible velocity fields compatible with homogeneous strain-rate boundary conditions, i.e.

$$\mathbf{v} = \mathbf{D}\mathbf{x}, \text{ for any } \mathbf{x} \in \partial\Omega. \tag{8}$$

In this paper, we will use this kinematic homogenization approach to obtain 3-D plastic potentials for porous solids with matrix obeying von Mises and Tresca yield criterion, respectively. The limit analysis is conducted for general 3-D states, i.e.

$$\mathbf{D} = D_1 \mathbf{e}_1 \otimes \mathbf{e}_1 + D_2 \mathbf{e}_2 \otimes \mathbf{e}_2 + D_3 \mathbf{e}_3 \otimes \mathbf{e}_3, \tag{9}$$

with $D_1, D_2, D_3$ being the eigenvalues (unordered) of $\mathbf{D}$ and ($\mathbf{e}_1, \mathbf{e}_2, \mathbf{e}_3$) its eigenvectors. We consider that the voids are spherical and randomly distributed in the matrix, so the porous material is isotropic. Therefore, the strain-rate potential of the porous solid, $\Pi(\mathbf{D}, f)$, is a scalar isotropic function. By the usual mathematical arguments based on theorems of representation of scalar isotropic functions (e.g. Boehler, 1987), it follows that $\Pi(\mathbf{D}, f)$ should depend on the strain-rate tensor $\mathbf{D}$ through its eigenvalues, or equivalently, through its invariants, which are: $D_m = (D_1 + D_2 + D_3)/3$, and the second and third-invariant of $\mathbf{D}'$, the deviator of $\mathbf{D}$, respectively, i.e.

$$\Pi(\mathbf{D}, f) = \Pi(D_m, J_{2D}, J_{3D}), \tag{10}$$

where $J_{2D} = \sqrt{(D_1'^2 + D_2'^2 + D_3'^2)/2}$ and $J_{3D} = D_1' D_2' D_3'$, with $D_i' = D_i - D_m$, i = 1...3. This choice of the representation of the SRP in terms of the invariants of $\mathbf{D}$ is based on physical grounds and is ultimately confirmed by experimental evidence (e.g. Spitzig et al., 1988, etc. ), the dependence of the first-invariant, $D_m$, in addition to the invariants of $\mathbf{D}'$, i.e. plastic dilatancy, being associated with the presence of voids.

In particular, in order to investigate the role played by the mean strain rate, $D_m$, on the overall mechanical response, for a fixed value of the porosity *f*, the shape of the cross-sections of the SRP with the deviatoric planes $D_m$ = constant, need to be determined. For this purpose, it is convenient to introduce the Oxyz frame of unit vectors ($\mathbf{e}_x, \mathbf{e}_y, \mathbf{e}_z$), which are related to the eigenvectors ($\mathbf{e}_1, \mathbf{e}_2, \mathbf{e}_3$) by the following relations:

$$\mathbf{e}_x = \frac{1}{\sqrt{3}}(\mathbf{e}_1 + \mathbf{e}_2 + \mathbf{e}_3), \quad \mathbf{e}_y = -\frac{1}{\sqrt{2}}(\mathbf{e}_1 - \mathbf{e}_2), \quad \mathbf{e}_z = \frac{1}{\sqrt{6}}(2\mathbf{e}_3 - \mathbf{e}_1 - \mathbf{e}_2). \tag{11}$$

Consider an arbitrary state represented by a point P ($D_1, D_2, D_3$) on the SRP isosurface $\Pi(\mathbf{D}, f)$ = constant. Since the Ox-axis coincides with the hydrostatic axis, the plane that contains the state P and is parallel to the Oyz-plane contains all the states belonging to the SRP with the same $D_m$. Thus, the intersection of the SRP with the deviatoric plane $D_m$ = constant is obtained by expressing the SRP in the (xyz) coordinates and then imposing $D_x$ = constant. Indeed, the SRP of a porous solid can be expressed as:

$$\Pi(\mathbf{D}, f) = \Pi(D_m, D'_1, D'_2, D'_3, f) = \Pi(D_x, D_y, D_z, f),$$

where

$$D_x = \sqrt{3}\, D_m$$
$$D_y = (D'_2 - D'_1)\frac{\sqrt{2}}{2} \qquad (12)$$
$$D_z = \sqrt{\frac{3}{2}}\, D'_3$$

Thus, any point P($D_1$, $D_2$, $D_3$) belonging to the intersection of the SRP locus with any deviatoric plane is characterized by two polar-type coordinates, $(R, \gamma)$ (see also Fig. 2(a)):

$$R = |OP| = \sqrt{D'^2_1 + D'^2_2 + D'^2_3} = \sqrt{2 J_{2D}}, \qquad (13a)$$

while $\gamma$ denotes the angle between $\mathbf{e}_y$ and **OP**, so

$$\tan(\gamma) = \frac{D_z}{D_y} = \sqrt{3}\,\frac{D'_3}{D'_2 - D'_1}. \qquad (13b)$$

Any state **D** belonging to the SRP surface is solely defined by: ($D_m$, R, $\gamma$).

Let $\mathbf{f}_i$ be the projections of the eigenvectors $\mathbf{e}_i$, i = 1…3 on a deviatoric plane. Obviously, $\mathbf{f}_3 = \mathbf{e}_z$ (see Eq. (11) and Fig. 2(a)). For a porous solid that is isotropic, the SRP has three-fold symmetry. Therefore, it is sufficient to determine the shape of the cross-section, i.e. $R = R(\gamma)$, only in the sector $-\pi/6 \leq \gamma \leq \pi/6$, the shape in all the other sectors being obtained by symmetry arguments. The sector $-\pi/6 \leq \gamma \leq \pi/6$, corresponds to the following ordering of the principal values of **D'**: $D'_2 \geq D'_3 \geq D'_1$. In particular, the sub-sector $-\pi/6 \leq \gamma \leq 0$ corresponds to states on the SRP for which ($D'_2 \geq 0$, $D'_3 \leq 0$, $D'_1 \leq 0$) so the third-invariant $J_{3D} > 0$ while the sub-sector $0 \leq \gamma \leq \pi/6$ corresponds to states for which ($D'_2 \geq 0$, $D'_3 \geq 0$, $D'_1 \leq 0$) so $J_{3D} < 0$ (see Eq. (13) and also Fig. 2(b)). Axisymmetric states correspond to either $\gamma = -\pi/6$ ($D'_1 = D'_3 < D'_2$) or $\gamma = \pi/6$ ($D'_2 = D'_3 > D'_1$). Note that for $-\pi/6 \leq \gamma \leq \pi/6$, $D'_1, D'_2, D'_3$ can be expressed as:

$$D'_1 = -\frac{R(\gamma)}{\sqrt{6}}\left(\sqrt{3}\cos\gamma + \sin\gamma\right)$$

$$D'_2 = \frac{R(\gamma)}{\sqrt{6}}\left(\sqrt{3}\cos\gamma - \sin\gamma\right) \quad (14)$$

$$D'_3 = \frac{2R(\gamma)}{\sqrt{6}}\sin\gamma$$

with $\sin 3\gamma = -\dfrac{27}{2}\cdot\dfrac{J_{3D}}{(J_{2D})^{3/2}}$ (see Eq. (13)).

The angle $\gamma$, which is a measure of the combined effects of the second and third-invariants, is related to the dimensionless parameter $\nu$ introduced by Drucker (1949),

$$\nu = \frac{D'_{int}}{D'_{min} - D'_{max}}, \quad (15)$$

where $D'_{min} = \min(D'_1, D'_2, D'_3)$, $D'_{max} = \max(D'_1, D'_2, D'_3)$ while $D'_{int}$ is the intermediate principal value.

## 3. Three-dimensional strain-rate potentials for porous solids with von Mises and Tresca matrices containing spherical voids

### 3.1. Preliminaries

For spherical void geometry an appropriate representative volume element (RVE) is a hollow sphere. Let $a$ denote its inner radius and $b = a f^{-1/3}$ the outer radius. The limit analysis is conducted for 3-D conditions for both tensile and compressive states. We use the trial velocity field $\mathbf{v}$, deduced by Rice and Tracey (1969), namely

$$\mathbf{v} = \mathbf{v}^v + \mathbf{v}^S, \quad (16)$$

where $\mathbf{v}^v$ describes the expansion of the cavity while $\mathbf{v}^S$ is associated to changes in the shape of the cavity. Imposing the boundary conditions and the constraint of matrix incompressibility, i.e. :

$$\mathbf{v}(\mathbf{x} = b\mathbf{e}_r) = \mathbf{D}\mathbf{x} \text{ and } \mathrm{div}(\mathbf{v}) = 0,$$

where $\mathbf{x}$ is the Cartesian position vector that denotes the current position in the RVE and $\mathbf{e}_r$ is the radial unit vector, it follows that:

$$\mathbf{v}^v = (b^3/r^2)D_m\mathbf{e}_r \text{ and } \mathbf{v}^S = \mathbf{D}'\mathbf{x}, \quad (17)$$

where $r = \sqrt{x_1^2 + x_2^2 + x_3^2}$ is the radial coordinate.

It is worth noting that if the plastic flow in the matrix is governed by either von Mises or Tresca criterion, the exact solution of the problem of a hollow sphere subjected to hydrostatic states (i.e. $\mathbf{D}' = 0$) is the same (see Lubliner, 2008). Furthermore, this exact solution is the term $\mathbf{v}^v = (b^3/r^2) D_m \mathbf{e}_r$ given by Eq. (17). Thus, for purely hydrostatic states, the mechanical response of a porous Tresca or a porous Mises solid is the same.

In the Cartesian basis ($\mathbf{e}_1$, $\mathbf{e}_2$, $\mathbf{e}_3$) associated with the eigenvectors of $\mathbf{D}$ (see Eq. (9)), the local strain rate tensor $\mathbf{d} = (\nabla \mathbf{v} + \nabla \mathbf{v}^T)/2$ corresponding to the velocity field given by Eq. (17) is :

$$d_{11} = D_1' + b^3 D_m \frac{1 - 3x_1^2/(x_1^2 + x_2^2 + x_3^2)}{(x_1^2 + x_2^2 + x_3^2)^{3/2}},$$

$$d_{22} = D_2' + b^3 D_m \frac{1 - 3x_2^2/(x_1^2 + x_2^2 + x_3^2)}{(x_1^2 + x_2^2 + x_3^2)^{3/2}}$$

$$d_{33} = D_3' + b^3 D_m \frac{1 - 3x_3^2/(x_1^2 + x_2^2 + x_3^2)}{(x_1^2 + x_2^2 + x_3^2)^{3/2}} \tag{18}$$

$$d_{12} = -\frac{3b^3 D_m x_1 x_2}{(x_1^2 + x_2^2 + x_3^2)^{3/2}}; \quad d_{13} = -\frac{3b^3 D_m x_1 x_3}{(x_1^2 + x_2^2 + x_3^2)^{3/2}}; \quad d_{23} = -\frac{3b^3 D_m x_2 x_3}{(x_1^2 + x_2^2 + x_3^2)^{3/2}}.$$

### 3.2. Three-dimensional strain-rate potential for a porous solid with von Mises matrix

Since the velocity $\mathbf{v}$ is incompressible and compatible with homogeneous strain rate boundary conditions (Eq. (17)), Hill-Mandel lemma applies. Thus, an upper-bound estimate of the exact plastic potential of porous solid with von Mises matrix is:

$$\Pi_{\text{Mises}}^+ (\mathbf{D}, f) = \frac{\sigma_T}{V} \int_\Omega \pi_{\text{Mises}} (\mathbf{d}) dV , \tag{19}$$

with $V = 4\pi b^3/3$, $\Omega$ is the domain occupied by the matrix, and $\pi_{Mises}(\mathbf{d})$ is the local plastic dissipation associated to the von Mises criterion (see Eq. (5)) for $\mathbf{d}$ given by Eq. (18), i.e.

$$\pi_{Mises}(\mathbf{d}) = \sigma_T \sqrt{(2/3)\left(D_1'^2 + D_2'^2 + D_3'^2\right) + 4 D_m^2 (b/r)^6 - 4 D_m (b/r)^3 \left(D_1' x_1^2 + D_2' x_2^2 + D_3' x_3^2\right)} \quad . \tag{20}$$

For general 3-D states, the integral given by Eq. (19) cannot be amenable to an exact analytic calculation. However, very recently Cazacu et al. (2013) have shown that for axisymmetric states the integrals expressing the SRP can be calculated explicitly, without making the approximations considered by Gurson (1975; 1977).

Given that the $\Pi_{Mises}^+(\mathbf{D}, f)$ is an even function of $\mathbf{D}$, only its expressions for the axisymmetric states corresponding to: ($D_m \geq 0$ and $D_1' = D_2' \geq 0$) and ($D_m \geq 0$ and $D_1' = D_2' \leq 0$) need to be calculated and will be given in the following; for all other axisymmetric states, the respective expressions are obtained by symmetry (see also Fig. 2(b)).

Let denote

$$u = \frac{|D_m|}{\max_{i=1..3}(|D_i'|)}. \tag{21}$$

(i) For $D_m \geq 0$ and $D_1' = D_2' \geq 0$:

$$\Pi_{Mises}^+(\mathbf{D}, f) = 2\sigma_T D_m \left[F(\sqrt{u/f}) - F(\sqrt{u})\right] \tag{22a}$$

with

$$F(z) = -2/(3z^2) + \frac{1}{3\sqrt{3}}\left[\tan^{-1}(2z + \sqrt{3}) - \tan^{-1}(2z - \sqrt{3})\right] +$$

$$+ \ln\sqrt{z^4 - z^2 + 1} + \frac{3z^4 + 3z^2 - 1}{6\sqrt{3}z^3} \ln\left(\frac{z^2 + z\sqrt{3} + 1}{z^2 - z\sqrt{3} + 1}\right).$$

(ii) For $D_m \geq 0$ and $D_1' = D_2' \leq 0$:

$$\begin{cases} \Pi_{Mises}^+(\mathbf{D}, f) = 2\sigma_T D_m \left[G(\sqrt{u/f}) - G(\sqrt{u})\right], \forall\, u < f \\ \Pi_{Mises}^+(\mathbf{D}, f) = 2\sigma_T D_m \left[G(\sqrt{u}) + G(\sqrt{u/f}) + 2\ln(3) - \frac{2}{9}\frac{\pi}{\sqrt{3}}\right], \forall\, f < u < 1 \\ \Pi_{Mises}^+(\mathbf{D}, f) = 2\sigma_T D_m \left[G(\sqrt{u}) - G(\sqrt{u/f})\right], \forall\, u > 1 \end{cases}$$

$$\tag{22b}$$

with:

$$G(z) = -2/(3z^2) - \frac{3z^4-3z^2-1}{3\sqrt{3}\,z^3}\tan^{-1}\left(\frac{z\sqrt{3}}{1-z^2}\right) + \frac{1}{3\sqrt{3}}\left(\tan^{-1}\left(\frac{2z+1}{\sqrt{3}}\right) - \tan^{-1}\left(\frac{2z-1}{\sqrt{3}}\right)\right) - \ln\sqrt{z^4+z^2+1}\ .$$

For all other loadings, $\Pi_{Mises}^{+}(\mathbf{D})$ (see Eq. (19)) cannot be calculated analytically and numerical integration methods need to be used. As already mentioned, it is sufficient to evaluate this SRP in the sector $-\pi/6 \le \gamma \le \pi/6$. Using Eq. (14), the integral expressing the SRP can be put in the form:

$$\Pi_{Mises}^{+}(\mathbf{D}, f) = \frac{\sigma_T}{V}\int_\Omega 2\sigma_T \sqrt{(R^2/6) + D_m^2 (b/r)^6 - 4\,R\,D_m (b/r)^3 F(\gamma,\, x_i^2)/(r^2/\sqrt{6})}\ dV \quad (23)$$

with $F(\gamma, x_1^2, x_2^2, x_3^2) = \sqrt{3}\,(x_2^2 - x_1^2)\cos\gamma + (2x_3^2 - x_1^2 - x_2^2)\sin\gamma$.

The integration is further simplified by making a change of coordinates from the coordinate system ($\mathbf{e}_1$, $\mathbf{e}_2$, $\mathbf{e}_3$) of the eigenvectors of $\mathbf{D}$ and Cartesian coordinates ($x_1$, $x_2$, $x_3$) to spherical coordinates. For axisymmetric loadings, the integral estimated numerically was compared to the exact result (Eq. (22)), differences being negligible (less than $10^{-7}$).

As an example, in Fig. 3(a) is shown a 3-D isosurface of the von Mises porous solid (calculated using Eq. (23)) corresponding to a porosity $f = 1\%$ for both tensile ($D_m = \text{tr}(\mathbf{D}) > 0$) and compressive ($D_m < 0$) states. Specifically, this convex surface contains all the points ($D_m$, R, $\gamma$) that produce the same plastic dissipation $\Pi_{Mises}^{+}(\mathbf{D}, f) = 9.21\cdot 10^{-3}$ for the porous solid. First, let us note that the presence of voids induces a strong influence of the mean strain rate $D_m$ on the plastic dissipation (see Fig. 3(b) which compares the isosurface for the fully-dense material (cylinder) with the isosurface for the porous solid ($f = 1\%$)). The SRP for $f = 1\%$ is closed on the hydrostatic axis. Indeed, for purely hydrostatic states (i.e. $\mathbf{D} = D_m^H \mathbf{I}$) according to Eq. (22), $\Pi_{Mises}^{+}(\mathbf{D}, f) = 2\sigma_T |D_m^H|\ln f$. For $f = 1\%$ and plastic dissipation of $9.21\cdot 10^{-3}$, this corresponds to: $D_m^H = \pm 1\cdot 10^{-3}\ \text{s}^{-1}$. Thus, the intersection of the isosurface with the planes $D_m = 1\cdot 10^{-3}\ \text{s}^{-1}$ and $D_m = -1\cdot 10^{-3}\ \text{s}^{-1}$, respectively are two points on the hydrostatic axis that are symmetric with respect to the origin (see also Fig. 3(b)).

To fully assess the effects of all invariants of the strain rate tensor, $\mathbf{D}$, on the plastic response of the porous solid, the cross-sections of the same 3-D isosurface with several deviatoric planes $D_m = $ constant are considered (see Fig.4). Note that the intersection with the plane $D_m = 0$, is a circle. This is to be expected since states for which $D_m = 0$ correspond to purely deviatoric loadings for which the plastic dissipation of the porous solid coincides with that of the matrix (von Mises behavior). The cross-sections with all the other deviatoric planes $D_m = $ constant, have three-fold symmetry with respect to the origin, and deviate slightly from a circle. This indicates that the third-invariant $J_{3D} = D'_1 D'_2 D'_3$ affects the plastic response of a porous solid with von Mises matrix. To assess the combined effects of $J_{3D}$ and $J_{2D}$, it is sufficient to study $R(\gamma)$ in the sector: $-\pi/6 \leq \gamma \leq \pi/6$ (i.e. study how the distance from the point on the cross-section and the origin evolves with $\gamma$). In this sector, axisymmetric conditions correspond to $\gamma = -\pi/6$ ($D_1 = D_3 < D_2$) or $\gamma = \pi/6$ ($D_2 = D_3 < D_1$). As an example, in Fig.5 is plotted $R(\gamma)$ (normalized by $R(\gamma = -\pi/6)$) for the cross-section corresponding to $D_m = 6 \cdot 10^{-4} \, s^{-1}$ and $D_m = 0$ (matrix behavior), respectively. As already mentioned, the cross-section $D_m = 0$ is a circle, so: $R(\gamma) = \text{constant} = R(\gamma = -\pi/6)$ (i.e. it is a straight line). As concerns the cross-section $D_m = 6 \, 10^{-4} \, s^{-1}$, note the influence of the third-invariant $J_{3D}$ (or $\gamma$) as evidenced by the deviation of $R(\gamma)/R(\gamma = -\pi/6)$ from a straight line. Note that the most pronounced difference is between the axisymmetric states, i.e. between $R(\gamma = -\pi/6)$ and $R(\gamma = \pi/6)$. The noteworthy result is that this holds true irrespective of the level of $D_m$ (see also Fig. 4) i.e. the shape of the cross-sections are similar and the most pronounced deviation from a circle is for axisymmetric states.

A remarkable property of the exact plastic potentials (stress-based and strain-rate based) of a porous solid with von Mises matrix is their centro-symmetry. This property is preserved by $\Pi^+_{\text{Mises}}(\mathbf{D}, f)$. This means that for any porosity $f$: $\Pi^+_{\text{Mises}}(D_m, R, \gamma) = \Pi^+_{\text{Mises}}(-D_m, R, -\gamma)$, i.e. the surface is symmetric with respect to the origin (see also Eq. (23) and Fig.3(a)). To further illustrate this noteworthy property, in Fig.6 are shown the cross-sections of the same 3-D isosurface $\Pi^+_{\text{Mises}}(\mathbf{D}, f) = 9.21 \cdot 10^{-3}$ ($f = 0.01$) with a deviatoric plane corresponding to a positive mean strain rate ($D_m = 6 \cdot 10^{-4} \, s^{-1}$, interrupted line) and a compressive mean strain rate ($D_m = -6 \cdot 10^{-4} \, s^{-1}$, solid line), respectively. The symmetry of the respective cross-sections with respect to the origin is clearly seen. For example, for loadings corresponding to $J_{3D} > 0$ (i.e. $-\pi/6 < \gamma < 0$) in order to produce the same plastic dissipation $R(\gamma)$ (or $\sqrt{2J_{2D}}$) must be higher for compressive states ($D_m < 0$ -interrupted line) than for tensile states ($D_m > 0$-solid line). The reverse holds true for loadings corresponding to $J_{3D} < 0$ ($0 < \gamma < \pi/6$).

More specifically, Fig. 6 clearly shows that for a given $D_m$, in order to reach the same plastic dissipation in the porous solid, there should be a very specific dependence between the invariants of $\mathbf{D}'$ i.e. between R and $\gamma$. Specifically, analysis of the cross-sections show that for $D_m > 0$, $R(\gamma)$ is a monotonically decreasing function of $\gamma$ (see also Fig. 5) while for $D_m < 0$, $R(\gamma)$ must be a monotonically increasing function of $\gamma$.

Comparison of the 3-D isosurfaces of the porous von Mises material with several deviatoric planes $D_m = \text{constant}$ for $f = 0.001$ and $f = 0.05$ are shown in Fig. 7(a) and Fig. 7(b), respectively. Irrespective of the level of porosity, the same general trends are observed. Namely, due to the presence of the voids, the shape of the cross-section changes very little from a circle as the absolute value of $D_m$ increases. However, the effect of the third-invariant becomes more pronounced with increasing porosity. For example, in the case when $f = 0.001$ the maximum deviation from a circle along all the cross-sections $D_m$ =constant is of 1.8%, whereas for a porosity $f = 0.05$, the maximum deviation is of 2.7%.

In summary, although the influence of the third-invariant is small it affects how plastic energy is accumulated in the porous solid and consequently how the porosity evolves. Furthermore, for a given $D_m$, the loading parameter $R(\gamma)$ is always between $R(\gamma = \pi/6)$ and $R(\gamma = -\pi/6)$, those limits corresponding to axisymmetric states. It follows that the most influence of the parameter $\gamma$ (or $J_{3D}$) on the dilatational response of the porous solid (consequently its influence on void growth or void collapse) occurs for these states. It is very worth noting that the same conclusions concerning the influence of the third-invariant on void evolution were drawn in their seminal study by Rice and Tracey (1969) for the case of large positive and negative triaxialities.

For metallic materials, the void volume fraction is generally small, so the volume of a sphere that circumscribes the cubical cell is extremely close to that of the cube. The main features of the macroscopic response predicted by the new criterion are confirmed by numerical calculations using cubic cells, namely the centro-symmetry of the surface and the coupling between the first invariant and the invariants of the deviator (see Cazacu et al., 2013; Alves et al., 2014 for comparison of porosity evolution according to the analytical model of Eq. (22) and FE model calculations of porosity evolution under axisymmetric loadings).

**3.3. Three-dimensional strain-rate potential for porous solids with Tresca matrix**

An upper-bound estimate of the overall plastic potential of porous solid with Tresca matrix is:

$$\Pi_{\text{Tresca}}^+ (\mathbf{D}, f) = \frac{\sigma_T}{V} \int_\Omega \pi_{\text{Tresca}} (\mathbf{d}) \, dV \ , \tag{24}$$

where $\pi_{\text{Tresca}}(\mathbf{d})$ is the local plastic dissipation associated to the Tresca criterion,

$$\pi_{\text{Tresca}}(\mathbf{d}) = \sigma_T \left( |d_1| + |d_2| + |d_3| \right) \tag{25}$$

where $d_1$, $d_2$, $d_3$ are the principal values (unordered) of the strain-rate field $\mathbf{d}$, given by Eq. (18). It is worth noting that given the symmetry properties of $\mathbf{d}$ (see Eq. (18)) and that $\pi(\mathbf{d})$ is an even function, $\Pi_{\text{Tresca}}^+ (\mathbf{D}, f)$ is also an even function, invariant under the transformation: $(D_m, \mathbf{D}') \to (-D_m, -\mathbf{D}')$. Thus, for any porosity, $f$, $\Pi_{\text{Tresca}}^+ (\mathbf{D}, f)$ is centro-symmetric (see also Fig. 8(a)).

Note that a major difficulty in obtaining a closed-form expression of this SRP is that $\pi_{\text{Tresca}}(\mathbf{d})$ depends on the sign of each of the principal values of the local strain-rate tensor, $\mathbf{d}$ (see Eq. (25)). This is a direct consequence of Tresca's yield criterion being dependent on the third-invariant of the stress deviator.

Only for axisymmetric loadings, the signs of the principal values $d_1$, $d_2$, $d_3$ of the local strain rate tensor **d**, given by Eq. (18), can be determined analytically. For these loadings, very recently Cazacu et al. (2014) showed that the integrals expressing the overall plastic dissipation $\Pi^+_{Tresca}(\mathbf{D}, f)$ could be calculated explicitly, without any approximation (for more details about the calculations, the reader is referred to Cazacu et al. (2014)).

Using the notation for u given by Eq. (21), the expression of the plastic dissipation of the porous solid with Tresca matrix for axisymmetric loadings are:

(i) For $D_m \geq 0$ and $D'_1 = D'_2 \geq 0$:

$$\begin{cases} \Pi^+_{Tresca}(\mathbf{D},f) = \dfrac{\sigma_T D_m}{8}\left(F_1(u/f) - F_1(u)\right), \forall\, u < f \\ \Pi^+_{Tresca}(\mathbf{D},f) = \dfrac{\sigma_T D_m}{8}\left(F_2(u/f) - F_1(u)\right), \forall\, f < u < 1 \\ \Pi^+_{Tresca}(\mathbf{D},f) = \dfrac{\sigma_T D_m}{8}\left(F_2(u/f) - F_2(u)\right), \forall\, u > 1 \end{cases} \quad (26\,a)$$

with

$$F_1(y) = 1 - 16\ln(2) - 6/y + \frac{(3y^2 + 8y^{3/2} + 6y - 1)}{y^{3/2}}\ln\left(\frac{\sqrt{y}+1}{1-\sqrt{y}}\right) + 16\ln(1-\sqrt{y})$$

$$F_2(y) = 1 - 16\ln(2) - 6/y + \frac{(3y^2 + 8y^{3/2} + 6y - 1)}{y^{3/2}}\ln\left(\frac{\sqrt{y}+1}{\sqrt{y}-1}\right) + 16\ln(\sqrt{y}-1).$$

For $D_m \geq 0$ and $D'_1 = D'_2 \leq 0$:

$$\begin{cases} \Pi^+_{Tresca}(\mathbf{D},f) = \dfrac{\sigma_T D_m}{8}\left(G_1(u/f) - G_1(u)\right), \forall\, u < f \\ \Pi^+_{Tresca}(\mathbf{D},f) = \dfrac{\sigma_T D_m}{8}\left(G_2(u/f) - G_1(u) - 12 - 16\ln(2)\right), \forall\, f < u < 1 \\ \Pi^+_{Tresca}(\mathbf{D},f) = \dfrac{\sigma_T D_m}{8}\left(G_2(u/f) - G_2(u)\right), \forall\, u > 1 \end{cases}$$

with:

$$G_1(y) = -6/y - \frac{3y^2 - 6y - 1}{y^{3/2}}\arctan\left(\frac{2\sqrt{y}}{y-1}\right) - 8\ln(y+1)$$

$$G_2(y) = 6/y + \frac{3y^2 - 6y - 1}{y^{3/2}}\arcsin\left(\frac{2\sqrt{y}}{y+1}\right) + 8\ln(y+1)$$

(26 b)

Due to the centro-symmetry of the porous Tresca SRP, for all other axisymmetric strain paths the expression of $\Pi_{\text{Tresca}}^{+}(\mathbf{D}, f)$ is obtained from Eq. (26) by symmetry (see Cazacu et al., 2014). It is very worth noting that for purely hydrostatic loadings (i.e. $\mathbf{D} = D_m \mathbf{I}$)  $\Pi_{\text{Tresca}}^{+}(\mathbf{D}, f) = \Pi_{\text{Mises}}^{+}(\mathbf{D}, f) = 2\sigma_T |D_m| \ln f$ (see Eq. (22) and Eq. (26), respectively).

For general 3-D states the overall plastic dissipation $\Pi_{\text{Tresca}}^{+}(\mathbf{D}, f)$ (Eq. (24)) can be estimated only numerically. For axisymmetric loadings, the numerical values are very close to the analytical ones obtained using Eq. (26) (error less than $10^{-7}$). As an example, in Fig.8(a) is shown a normalized ($\sigma_T = 1$) 3-D isosurface of the porous Tresca solid corresponding to a porosity $f = 1\%$ for states characterized by ($D_m > 0$) and ($D_m < 0$), respectively. Specifically, this convex surface contains all states ($D_m$, R, γ) that produce the same plastic dissipation $\Pi_{\text{Tresca}}^{+}(\mathbf{D}, f) = 9.21 \cdot 10^{-3}$ for the porous solid. The presence of voids induces a strong influence of the mean strain rate $D_m$ on the plastic dissipation (compare the isosurface for the porous solid with that for the fully dense material (hexagonal prism) shown in Fig.8(b)). The SRP for $f = 1\%$ is closed on the hydrostatic axis. As already mentioned for purely hydrostatic states, the SRP of a porous solid with Tresca matrix coincides with the SRP of a porous solid with von Mises matrix. Thus, for $f = 1\%$ and plastic dissipation of $9.21 \cdot 10^{-3}$, the intersection of the Tresca isosurface with the hydrostatic axis is also at: $D_m^H = \pm 1 \cdot 10^{-3} \text{ s}^{-1}$ (see also Fig. 8(a)).

To investigate the effects of all invariants on the response of the porous Tresca solid, the cross-sections of the same 3-D isosurface with deviatoric planes $D_m$ = constant are considered (see Fig. 9). Note that the intersection with the plane $D_m = 0$ is a regular hexagon. This is to be expected since states for which $D_m = 0$ correspond to purely deviatoric loadings for which the plastic dissipation of the porous solid coincides with that of the matrix (i.e. Tresca behavior, see also Fig.1(a)).

It is very interesting to note the very strong influence of $D_m$ on the plastic behavior of the porous Tresca material, the shape of the cross-sections changing drastically with the level of $D_m$ = constant. Due to the presence of voids, all cross-sections are smoothed out, their shape evolving from a hexagon ($D_m = 0$) to a triangle with rounded corners (the innermost cross-section corresponding to $D_m = 9 \cdot 10^{-4} \text{ s}^{-1}$).

It is also worth noting the strong coupling between $D_m$, R, (or $J_{2D}$), and γ (measure of $J_{3D}$ and $J_{2D}$, see Eq. (13)). Note that the evolution of R with γ is very specific and depends strongly on $D_m$ as shown in the zoom of the same cross-sections in the sector: $-\pi/6 \leq \gamma \leq \pi/6$ (see Fig. 9(b)). To better assess the importance of this coupling between all invariants, in Fig. 10 is plotted $R(\gamma)$ (normalized by $R(\gamma = -\pi/6)$) for cross-sections corresponding to $D_m$ = constant ( in the range $D_m = 0$ to $D_m = 9 \cdot 10^{-4}$). Since Tresca's criterion depends on both $J_{2D}$ and $J_{3D}$, even the cross-section corresponding to $D_m = 0$ is not a circle. Equivalently, compare Fig 10 depicting $R(\gamma)/R(\gamma = -\pi/6)$ vs. $\gamma$ at $D_m = 0$ for the porous Tresca solid with the evolution corresponding to $D_m = 0$ in the case of a von Mises matrix (i.e. the straight line in Fig. 5). Furthermore, the analysis of the evolution of $R(\gamma)$ at $D_m = 0$ for a porous Tresca material shows that it has a maximum at $\gamma = 0$ (i.e. states corresponding to $J_{3D} = 0$) while the minima correspond to the axisymmetric states, for which $\gamma = -\pi/6$ and $\gamma = \pi/6$, respectively. Note also that only for $D_m = 0$ (i.e. matrix behavior) $R(\gamma = -\pi/6) = R(\gamma = \pi/6)$, i.e. the plastic dissipation is the same for the axisymmetric state corresponding to $J_{3D} > 0$ and the axisymmetric state corresponding to $J_{3D} < 0$, respectively.

Another noteworthy property is the strong influence of $D_m$ on the variation of R with $\gamma$. Specifically, with increasing $D_m$ the maximum of $R(\gamma)$ is no longer at $\gamma = 0$, but shifts toward the axisymmetric case corresponding to $\gamma = -\pi/6$ ($D_1 = D_3 < D_2$ and $J_{3D} > 0$); on the other hand, the minimum of $R(\gamma)$ is always obtained for $\gamma = \pi/6$ (axisymmetric state corresponding to $J_{3D} < 0$). Another specificity of the dilatational response of a porous Tresca solid is that irrespective of the cross-section $D_m$ = constant, there are two states with the same R: the axisymmetric state $\gamma = -\pi/6$ and another state say $\gamma = \gamma_1$; the value of $\gamma_1$ depends on $D_m$ (e.g. for $D_m = 0$, $\gamma_1 = \pi/6$, the higher $D_m$ the lower is $\gamma_1$).

As already mentioned, $\Pi^+_{Tresca}(\mathbf{D}, f)$ is centro-symmetric. This means that for any porosity $f$: $\Pi^+_{Tresca}(D_m, R, \gamma, f) = \Pi^+_{Tresca}(-D_m, R, -\gamma, f)$. To illustrate this remarkable property, in Fig. 11 are shown the cross-sections of the same 3-D isosurface $\Pi^+_{Tresca}(\mathbf{D}, f) = 9.21 \cdot 10^{-3}$ ($f = 0.01$) with the deviatoric planes $D_m = 7 \cdot 10^{-4}$ s$^{-1}$ and $D_m = 9 \cdot 10^{-4}$ s$^{-1}$, respectively (interrupted lines) as well as the cross-sections with the planes $D_m = -7 \cdot 10^{-4}$ s$^{-1}$ and $D_m = -9 \cdot 10^{-4}$ s$^{-1}$, respectively (solid lines). The symmetry of all cross-sections with respect to the origin is clearly seen. For example, for states corresponding to $J_{3D} > 0$ ( $-\pi/6 < \gamma < 0$ ) to produce the same plastic dissipation, R (or $J_{2D}$) must be higher for tensile states ($D_m > 0$ interrupted line) than for compressive states ($D_m < 0$: solid line). The reverse holds true for loadings corresponding to $J_{3D} < 0$ ($0 < \gamma < \pi/6$).

It is also very interesting to compare the behavior of porous solids with Mises and Tresca matrix, respectively. For a porous Tresca solid the shapes of the cross-sections with deviatoric planes are strongly dependent on the level of $D_m$, whereas for a porous Mises the shape of the cross-sections is similar irrespective of the level of $D_m$ (compare Fig. 4 with Fig. 9(a)). Let us also compare Fig. 10 (porous Tresca) and Fig. 5 (porous Mises), respectively. For a porous Tresca, the influence of $\gamma$ is very strong, Fig. 10 showing that the coupling between R and $\gamma$ (i.e. between $J_{2D}$ and $J_{3D}$) depends on the level of $D_m$. While in the case of the porous Mises, the most pronounced difference in plastic response is between the axisymmetric states (i.e. between $R(\gamma = \pi/6)$ and $R(\gamma = -\pi/6)$, for a porous Tresca no general conclusions can be drawn because the specific dependency of R with $\gamma$ (i.e. coupling between $J_{2D}$ and $J_{3D}$) depends both on the level of porosity and $D_m$ (see also Fig. 12-13).

To further illustrate the specificities of the plastic response of a porous solid with Tresca matrix in Fig 12 (a) are shown the cross-sections with deviatoric planes of the surface $\Pi_{Tresca}^{+}(\mathbf{D}, f = 0.001) = 9.21 \cdot 10^{-3}$ while in Fig. 12 (b)) are shown the cross-section of the surface $\Pi_{Tresca}^{+}(\mathbf{D}, f = 0.05) = 9.21 \cdot 10^{-3}$. Note that these surfaces correspond to the same value of the plastic dissipation ($9.21 \cdot 10^{-3}$), but to different porosities. Note that irrespective of the level of porosity, the same general trends are observed. Namely, due to the presence of the voids, the shape of the cross-section changes from a regular hexagon ($D_m = 0$; matrix behavior) to a triangle with rounded corners as the absolute value of $D_m$ increases. It is very interesting to note the importance played by the level of porosity. Indeed, the porosity is key in how fast the shape of the cross-section changes along the hydrostatic axis. If the level of porosity is small (e.g. $f = 0.001$ see Fig. 12(a)), the cross-sections smooth out slower than in the case when the level of porosity in the matrix is higher (compare with the cross-sections corresponding to $f = 0.01$ shown in Fig. 9). The same conclusion can be drawn by comparing the cross-sections shown in Fig. 9 ($f = 0.01$) and those presented in Fig. 12(b), which correspond to a porosity $f = 0.05$. For example, the cross-section of triangular shape corresponds to a lower value of $D_m/D_m^H$ in the case when $f = 0.05$ than in the case when $f = 0.01$.

## 4. Discussion on the importance of the plastic flow of the matrix on the dilatational response

The most widely used plastic potential for isotropic porous solids containing randomly distributed spherical voids was proposed by Gurson (1977). This yield criterion was derived by conducting limit analysis on a hollow sphere made of a rigid-plastic material obeying von Mises yield criterion using the trial velocity field deduced by Rice and Tracey (1969) (see Eq. (17)). In his analysis, Gurson (1975; 1977) assumed that the coupled effects between the mean strain rate $D_m$ and $\mathbf{D}'$ (i.e. the cross-term $D_m(b/r)^3 \left(D'_1 x_1^2 + D'_2 x_2^2 + D'_3 x_3^2\right)$ in the expression of $\pi_{Mises}(\mathbf{d})$ given by Eq. (20)) can be neglected, i.e.

$$\pi_{Mises}(\mathbf{d}) \cong \sigma_T \sqrt{4 D_m^2 (b/r)^6 + (2/3) R^2}$$

With this approximation, the SRP of a porous von Mises solid takes the following expression :

$$\Pi_{Gurson}(\mathbf{D}, f) = 2|D_m| \left[ \frac{\sqrt{1 + 6 D_m^2 / R^2} - \sqrt{f^2 + 6 D_m^2 / R^2}}{(D_m/R)\sqrt{6}} + \ln\left(\frac{1}{f} \cdot \frac{(D_m/R)\sqrt{6} + \sqrt{f^2 + 6 D_m^2 / R^2}}{(D_m/R)\sqrt{6} + \sqrt{1 + 6 D_m^2 / R^2}}\right) \right]$$

(27)

Note that $\Pi_{\text{Gurson}}(\mathbf{D},f)$ is the exact dual of the classic stress-based potential of Gurson (1977).

For axisymmetric states, Cazacu et al. (2014) have shown that if the same simplifying hypothesis considered by Gurson (1975; 1977) is made when evaluating the local plastic dissipation associated to Tresca's yield criterion (see Eq. (25)), then the truncated expression of the overall plastic dissipation at which one arrives coincides with $\Pi_{\text{Gurson}}(\mathbf{D},f)$, given by Eq. (27). This means that neglecting couplings between the mean strain rate and $\mathbf{D}'$ (i.e. couplings between normal and shear effects) the specificities of the plastic flow of the matrix are erased. It is worth comparing the strain-rate potential for a porous Mises material obtained by Gurson (1975) (i.e. Eq. (27)) with the exact 3-D strain-rate potential for a porous Mises material given by Eq. (19), and the strain-rate potential for a porous Tresca material given by Eq. (24 ). Note that:

- Only for three states, all three strain-rate potentials coincide. These states are: purely hydrostatic loading ( $\mathbf{D}' = 0$); and axisymmetric purely deviatoric states (two principal values of $\mathbf{D}'$ equal and $D_m = 0$).

- In contrast to the porous von Mises SRP ( Eq. (19) ), Gurson's SRP does not involve any dependence on $J_{3D}$ (see Eq. (27)). Therefore, the cross-section of Gurson's SRP with any deviatoric plane (i.e. irrespective of $D_m$ or the porosity level) is always a circle. ( R = constant) (see also Fig. 13).

- The porous von Mises SRP ( Eq. (19)) and the porous Tresca SRP (Eq. (25)) are centro-symmetric i.e. they are invariant to the transformation ( $D_m$, R, γ) → ( -$D_m$, R,-γ) (see Fig. 12). However, Gurson's SRP involves only $D_m$ and $J_{2D}$ (or R) and displays stronger symmetry properties, being also invariant to the transformation: ($D_m$, R) → (-$D_m$, -R). In other words, the dilatational response according to Gurson's SRP is insensitive to the sign of the mean strain-rate $D_m$.

The specific differences between the three SRPs are further illustrated in Fig. 13, which shows the cross-sections with several deviatoric planes of the respective isosurfaces corresponding to the same void volume fraction ( $f = 1\%$). Each of these surfaces corresponds to the same value of the plastic dissipation i.e. in Fig 13 are represented: $\Pi_{\text{Tresca}}^{+}(\mathbf{D}, f = 0.01) = 9.21 \cdot 10^{-3}$, $\Pi_{\text{Mises}}^{+}(\mathbf{D}, f = 0.01) = 9.21 \cdot 10^{-3}$, $\Pi_{\text{Gurson}}^{+}(\mathbf{D}, f = 0.01) = 9.21 \cdot 10^{-3}$, respectively. Since Gurson's SRP was obtained by truncating the overall plastic dissipation (see Eq. (27)), it is necessarily interior to the exact SRP, which is $\Pi_{\text{Mises}}^{+}$ (given by Eq. (19)). Irrespective of the level of $D_m$, Tresca's SRP is exterior to the surfaces corresponding to von Mises matrix. This means that Gurson's SRP is the most dissipative of the three SRP's, since in order to reach the same value of the plastic dissipation, the norm of the loading, R(γ), is lower than for a porous Mises (Eq. (19)) or porous Tresca (Eq. (24)). On the other hand, Tresca's SRP is the least dissipative potential.

The noteworthy result revealed is the very strong influence of the plastic flow of the matrix on the response of a porous solid. If the matrix obeys the von Mises criterion the shape of the cross-sections of the porous solid changes very little as $D_m$ increases, but if the matrix behavior is described by Tresca's criterion the shape of the cross-section evolves from a hexagon to a triangle with rounded corners (compare Fig. 13(a) which shows the cross-sections with the plane $D_m = 2 \cdot 10^{-4} \, s^{-1}$ with Fig. 13 (c) which shows the cross-section with the plane $D_m = 8 \cdot 10^{-4} \, s^{-1}$). Although the difference between the response of a porous solid with von Mises matrix and that with a Tresca matrix becomes less important with increasing $D_m$, it strongly affects void evolution (see the estimate of the differences in the rates of void growth for very high hydrostatic loadings reported by Rice and Tracey, 1969).

Comparison between the cross-sections of the 3-D isosurfaces of the porous Tresca material, porous Mises material, and Gurson's isosurface at γ = 0 and γ = -π/6 are shown in Fig. 14 (a) and Fig. 14 (b), respectively ($f = 0.01$). Due to the presence of voids, the SRP of the Tresca porous material is smooth, the normal to the surface being unique irrespective of the state.
Note also the strong effect of $J_{3D}$ ( or γ) on the shape of the cross-sections for the porous Tresca material. Only for three states, all strain-rate potentials coincide. These states are: purely hydrostatic loading ( R = 0); and axisymmetric purely deviatoric states (γ = -π/6 and $D_m$= 0 shown in Fig. 14(b) and for γ = π/6 and $D_m$= 0, see Eq. (19), (24), (27)).

5. Summary and conclusions

The aim of this paper was to investigate the properties of the 3-D plastic potentials for porous solids with Tresca and von Mises, matrices respectively. For the first time, the role of the plastic flow of the matrix on the dilatational response was analyzed for general 3-D conditions for both compressive and tensile states.

It has been shown that if the matrix is described by the von Mises criterion:

- The plastic response of the porous solid depends on all three invariants of the strain-rate tensor **D**.
- The corresponding 3-D surface is smooth and symmetric with respect to the origin.

- Its cross-sections with the deviatoric planes $D_m$ = constant ($\neq 0$) have three-fold symmetry with respect to the origin, and deviate slightly from a circle.

- The coupling between the invariants of **D'** i.e. between $R(=\sqrt{2J_{2D}})$ and $\gamma$ (measure of $\sqrt{J_{2D}}$ and $J_{3D}$) is very specific :

    - for $D_m > 0$, $R(\gamma)$ is a monotonically decreasing function of $\gamma$
    - for $D_m < 0$, $R(\gamma)$ is a monotonically increasing function of $\gamma$

- The strongest effect of the third-invariant is for axisymmetric states i.e. between $R(\gamma = \pi/6)$ and $R(\gamma = -\pi/6)$.

Therefore, the most influence of the parameter $\gamma$ (or $J_{3D}$) on void growth or void collapse occurs for axisymmetric states. It is very worth noting that the same conclusions concerning the influence of the third-invariant on void evolution were drawn by Rice and Tracey (1969) for the case of large hydrostatic stresses.

It has been shown that if the plastic behavior of the matrix is described by Tresca's criterion:

- The 3-D surface is centro-symmetric and displays a strong coupling between all invariants.

- The shapes of the cross-sections with deviatoric planes are strongly dependent on the level of $D_m$. As the absolute value of $D_m$ increases, the shape changes from a regular hexagon ($D_m = 0$) to a triangle with rounded corners.

- The level of porosity is key in how fast the shape changes with the mean strain rate $D_m$. If the level of porosity is small, the cross-sections smooth out slower than in the case when the level of porosity in the matrix is higher.

- For $D_m = 0$ (i.e. Tresca behavior) the maximum of $R(\gamma)$ is at $\gamma = 0$ ($J_{3D} = 0$), the minima being for axisymmetric states.

    For $D_m > 0$: the maximum of $R(\gamma)$ is not at $\gamma = 0$ anymore, but shifts toward the axisymmetric case corresponding to $\gamma = -\pi/6$ ($D_1 = D_3 < D_2$ and $J_{3D} > 0$); on the

other hand, the minimum of R(γ) is always obtained for γ = π/6 (axisymmetric state corresponding to $J_{3D} < 0$). The reverse holds true for $D_m < 0$.

- While in the case of the porous Mises solid, the most pronounced difference in the response is between the axisymmetric states (i.e. between R(γ = π/6) and R(γ = -π/6), for a porous Tresca no general conclusions can be drawn because the specific expression of R(γ) depends both on the level of porosity and the level of $D_m$.

The strain-rate potentials for porous solids with Tresca and von Mises matrices obtained in this work were compared to the exact conjugate in the strain-rate space of the classic Gurson (1977) stress-based potential. It was shown that only for three states, all three strain-rate potentials coincide. These states are: purely hydrostatic loading (**D'** = 0); and axisymmetric purely deviatoric states (two principal values of **D** equal and $D_m$= 0). In contrast to the exact porous von Mises SRP (Eq. (19)), the Gurson's SRP does not involve any dependence on $J_{3D}$. Therefore, its cross-sections with any deviatoric plane (irrespective of $D_m$ or the porosity level) is always a circle. The exact porous von Mises SRP and the porous Tresca SRP are centro-symmetric i.e. they are invariant to the transformation ($D_m$, R, γ) → (-$D_m$, R,-γ) (see Fig. 6 and Fig. 9, respectively). Gurson's SRP involves only dependence of $D_m$ and $J_{2D}$ (or R) is insensitive to the sign of the mean strain rate (compressive or tensile states). Most importantly, irrespective of the level of $D_m$, Gurson's SRP is the most dissipative of the three SRP's, since in order to reach the same value of the plastic dissipation, the norm of the loading, R(γ), is lower than that of the exact porous Mises or of the porous Tresca SRP. On the other hand, Tresca's SRP is the least dissipative potential.

The results presented in this paper provide fundamental understanding of the influence of all invariants on the response of porous solids ranging from polycrystalline metals to metallic foams for fully 3-D loadings. In particular, the porous Tresca model accounts for all the trends observed experimentally by Combaz et al. (2011) for aluminum foams, specifically the triangular shape of the surface and the strong influence of the sign of mean stress on the dilatational response (i.e. its centro-symmetry).

Since Tresca's criterion is the isotropic form of the yield criterion of a single crystal obeying Schmid law, the insights provided in this paper into the couplings between invariants on the response of a porous material with matrix having a faceted yield surface pave the way towards improving the understanding and further the modeling of the influence of porosity in single crystals subject to multi-axial loadings.

The importance of the coupling between the sign of mean stress and $J_3^\Sigma$ predicted by the new criteria need to be further assessed by numerical studies and experiments on damage evolution and fracture in engineering materials.

# References

Alves, J.L., Revil-Baudard, B., Cazacu, O., 2014. Importance of the coupling between the sign of the mean stress and the third-invariant on the rate of void growth and collapse in porous solids with von Mises matrix. Modelling and Simulation in Materials Science and Engineering, 22(2), 025005 (18 pp.).

Barlat, F., Chung, K., Richmond, O., 1993. Strain-rate potential for metals and its application to minimum plastic work path calculations, Int. J. Plasticity 9, 51–63.

Barlat, F., Brem, J.C., Yoon, J.W., Chung, K., Dick, R.E., Lege, D.J., Pourboghrat, F., Choi, S.-H., Chu, E., 2003. Plane stress yield function for aluminum alloy sheet-Part I: Theory. Int. J. Plasticity 19, 1297–1319.

Barsoum, I. and Faleskog, J., 2007. Rupture in combined tension and shear: Experiments. Int. J. Solids Struct., 44, 1768-1786.

Boehler, J.P.,1987. Application of tensors functions in solids mechanics. CISM courses and lectures. In: International Center for Mechanical Sciences, vol. 292, Springer-Verlag, Wien, New York.

Cazacu, O., Plunkett, B., Barlat, F., 2006. Orthotropic yield criterion for hexagonal closed packed metals. Int. J. Plasticity 22, 1171–1194.

Cazacu, O. and Stewart, J.B., 2009. Plastic potentials for porous aggregates with the matrix exhibiting tension-compression asymmetry. J. Mech. Phys. Solids, 57, 325-341.

Cazacu, O., Ionescu, I.R., Yoon, J.W., 2010. Orthotropic strain-rate potential for the description of anisotropy in tension and compression of metals. Int. J. Plasticity, 26, 887–904.

Cazacu, O., Revil-Baudard, B., Lebensohn, R. A., Garajeu, M., 2013. New analytic criterion describing the combined effect of pressure and third invariant on yielding of porous solids with von Mises matrix. J. Appl. Mech. 80 (6), 64501-1- 64501-5.

Cazacu, O., Revil-Baudard, B., Chandola N., Kondo, D., 2014. New analytical criterion for porous solids with Tresca matrix under axisymmetric loadings. . Int. J. Solids Struct. 51, 861-874.

Chung, K., Barlat, F., Brem, J.C., Lege, D.J., Richmond, O., 1997. Blank shape design for a planar anisotropic sheet based on ideal forming design theory and FEM analysis, Int. J. Mech. Sciences 39, 105–120.

Combaz, E., Bacciarini, C., Charvet, R., Dufour, W.,. Mortensen, A., 2011. Multiaxial yield behavior of Al replicated foam. J. Mech. Phys. Solids 59, 1777-1793.

Drucker, D. 1949. Relation of experiments to mathematical theories of plasticity. J. Appl. Mech. 16, 349–357.

Gologanu, M., 1997. Etude de quelques problemes de rupture ductile des metaux., Ph.D. Thesis, Univ. Pierre and Marie Curie, Paris, France.

Gurson, A., 1975. "Plastic flow and fracture behavior of ductile materials incorporating void nucleation, growth, and interaction". PhD Thesis, Brown University, Rhode Island.

Gurson, A. L., 1977. Continuum theory of ductile rupture by void nucleation and growth. Part I: Yield criteria and flow rules for porous ductile media. J. Engng. Matl. Tech. Trans. ASME, Series H, 99, 2-15.

Hill, R., 1948. A theory of yielding and plastic flow of anisotropic metals. Proc. Roy. Soc. London A 193, 281-297.

Hill, R., 1967. The essential structure of constitutive laws for metal composites and polycrystals. J. Mech. Phys. Solids 15, 79–95.

Hill, R., 1987. Constitutive dual potentials in classical plasticity. J. Mech. Phys. Solids 35, 23-33.

Khan, A. and Liu, H., 2012. A new approach for ductile fracture prediction on Al 2024-T351 alloy. Int. J. Plasticity, 35, 1-12.

Kim, D., Barlat, F., Bouvier, S., Raballah, M., Balan, T., Chung, K., 2007. Non-quadratic anisotropic potentials based on linear transformation of plastic strain rate. Int. J. Plasticity 23, 1380-1399.

Lebensohn, R.A., Cazacu, O., 2012. Effect of single-crystal plastic deformation mechanisms on the dilatational plastic response of porous polycrystals. Int. J. Solids Struct., 49, 3838-3852.

Lecarme, L., Tekoglu, C., Pardoen, T., 2011. Void growth and coalescence in ductile solids with stage III and stage IV strain hardening. Int. J. Plasticity 27, 1203-1223.

Lubliner, J. 2008. Plasticity theory, Dover Publications Inc., Mineola, New York.

Mandel, J., 1972. Plasticite classique et viscoplasticite, Int. Centre Mech Sci., Courses and lectures, 97, Udine 1971, Springer, Wien, New York.

Maire, E., Zhou, S., Adrien, J., Dimichiel, M., 2011. Damage quantification in aluminum alloys in situ tensile tests in X-ray tomography. Engineering Fracture Mechanics, 78, 2679-2690.

Richelsen, A.B., Tvergaard, V., 1994. Dilatant plasticity or upper bound estimates for porous ductile solids, Acta Metall. Mater. 42, 2561-2577.

Rice, J.R., Tracey, D.M., 1969. On the ductile enlargement of voids in triaxial stress fields. J. Mech. Phys. Solids 17, 201-217.

Salencon, J. 1983. Yield design and limit analysis. Eds. Presses de l'ENPC, Paris.

Spitzig, W.A., Smelser, R.E., Richmond, O., 1988. The evolution of damage and fracture in iron compacts with various initial porosities. Acta Metall. 36, 1201-1211.

Stoughton, T B. and Yoon, J. W., 2011, A new approach for failure criterion for sheet metals. Int. J. of Plasticity 27, 440–459.

Talbot, D. R. S. and Willis, J. R., 1985. Variational principles for inhomogeneous non-linear media, IMA J Applied Mathematics, 35(1), 39-54.

Tvergaard, V., 1981. Influence of voids on shear band instabilities under plane strain conditions. Int. J. Fracture 17, 389-407.

Tvergaard, V. and Nielsen, K.L. , 2010. Relation between a micro-mechanical model and a damage model for ductile failure in shear. J. Mech. Phys. Solids.., 58, 1243-1252.

Van Houtte, P. and Van Bael, A., 2004.  Convex plastic potentials of fourth and sixth rank for anisotropic materials. Int. J. Plasticity, 20, 1505–1524.

Yoon, J-H., Cazacu, O. and Yoon, J.W., 2011. Strain rate potential based elastic/plastic anisotropic model for metals displaying tension-compression asymmetry. Computer Methods in Applied Mecahnics and Engineering, 200,1993-2004.

Yoon, J.W., Lou, Y., Yoon, J-H, Glazov, M.V., 2013. Asymmetric yield functions based on stress invariant for  pressure-sensitive materials. Int. J. Plasticity,
http://dx.doi.org/10.1016/j.ijplas.2013.11.008

Ziegler, H., 1977. An introduction to thermodynamics, North-Holland, Amsterdam.

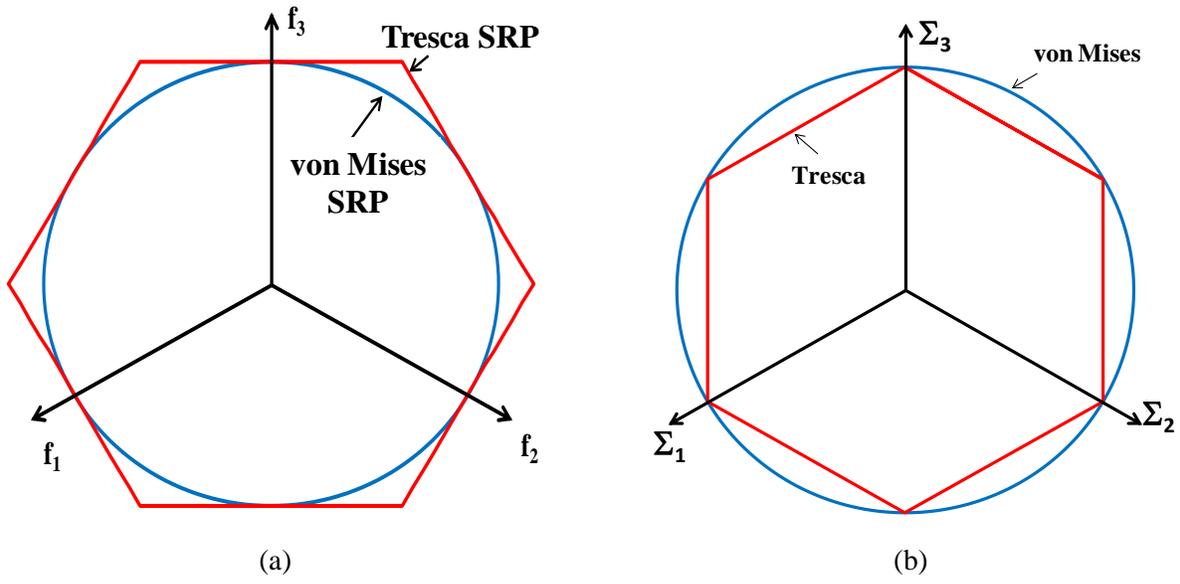

**Fig. 1**. (a) Section of the von Mises strain-rate potential (SRP) (Eq. (5)) and Tresca's SRP (Eq. (6)) with the octahedral plane; (b) representation of their respective duals in the stress space, i.e. the normalized von Mises and Tresca yield surfaces, respectively.

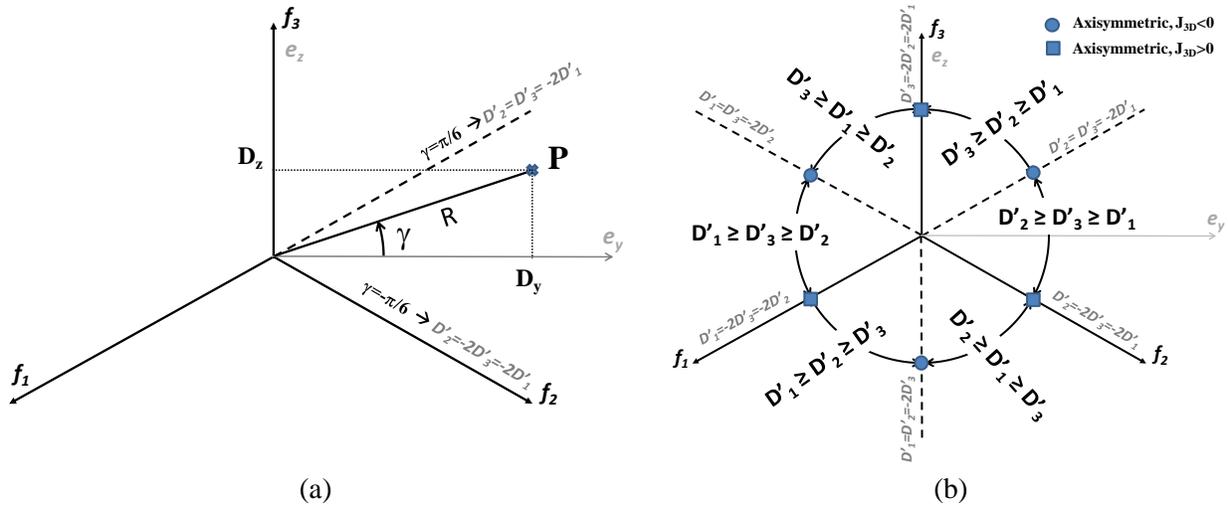

**Fig. 2**. (a) Definition of the polar-type coordinates $(R,\gamma)$, representing any state P ($D_1$, $D_2$, $D_3$) belonging to the intersection of a strain-rate potential isosurface with any deviatoric plane (plane of normal the hydrostatic axis) (b) General symmetry properties of the cross-section of the SRP for an isotropic material.

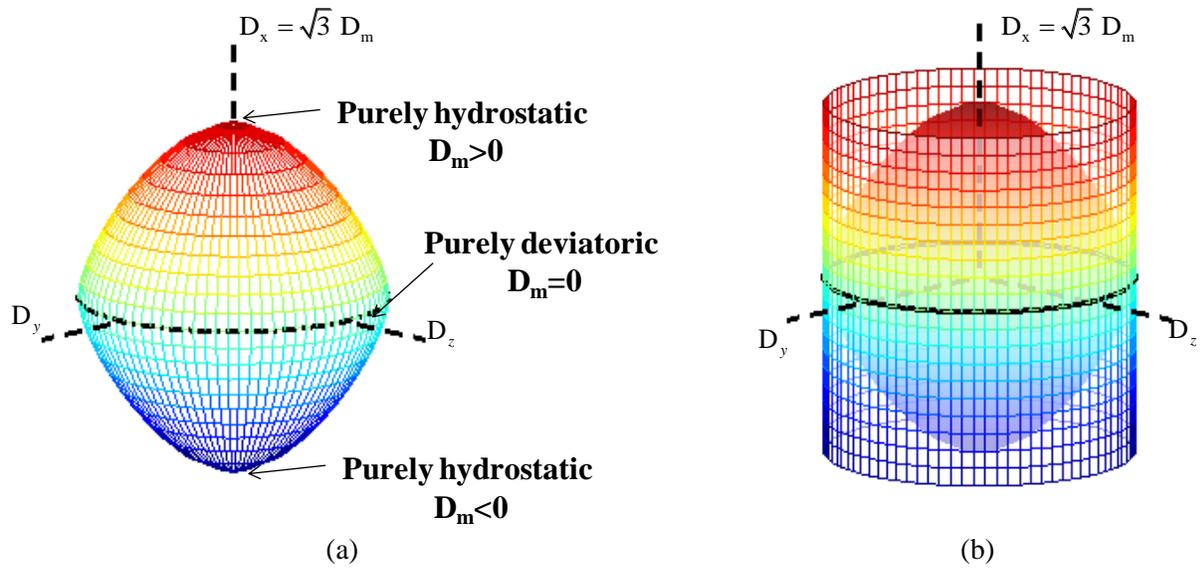

**Fig. 3**. (a) The 3-D surface for a porous solid with von Mises matrix according to Eq. (19) for both tensile mean strain rate ($D_m = \mathrm{tr}(\mathbf{D}) > 0$) and compressive ($D_m < 0$) states. Note that this convex surface contains all the points ($D_m$, $R$, $\gamma$) that produce the same plastic dissipation $\Pi^+_{\mathrm{Mises}}(\mathbf{D}, f) = 9.21 \cdot 10^{-3}$ for the porous solid. Initial porosity: $f = 0.01$. (b) Comparison between the 3-D isosurface of the porous material ($f = 0.01$) and the 3-D isosurface of the fully-dense material ($f = 0$: von Mises matrix behavior).

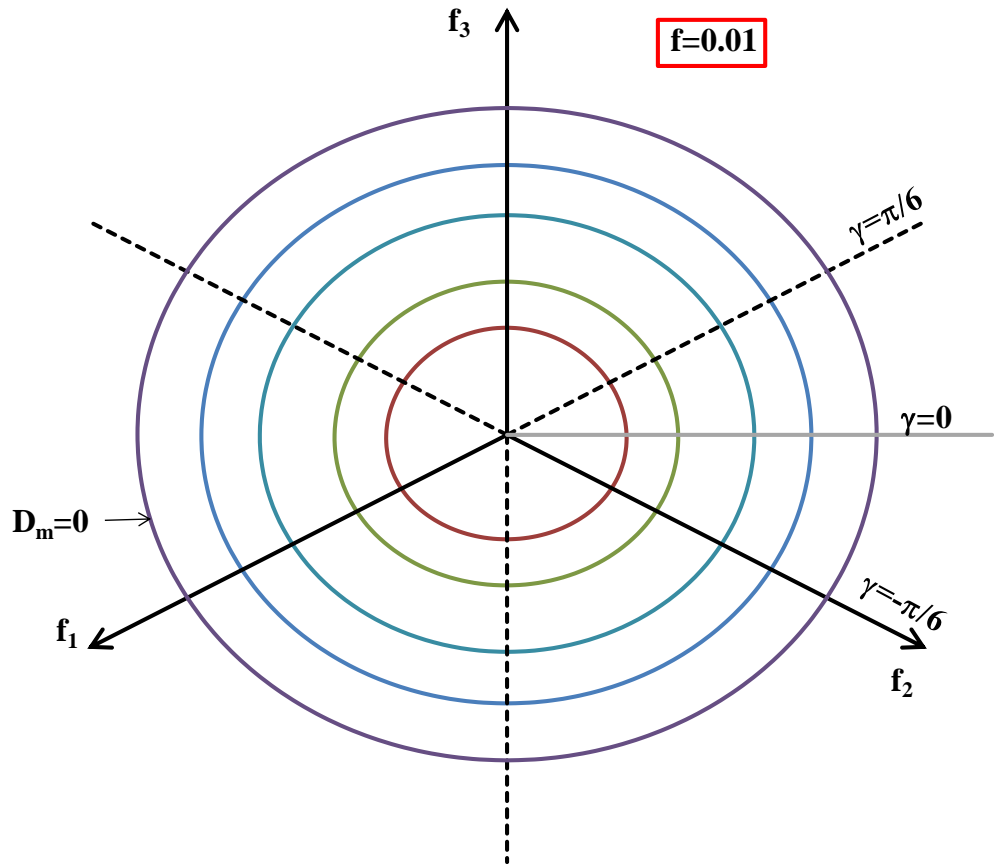

**Fig. 4**. Cross-sections of the 3-D isosurface of a porous von Mises material with several deviatoric planes $D_m$ = constant: outer cross-section represents the intersection with the plane $D_m = 0$ while the inner cross-section corresponds to $D_m = 9 \cdot 10^{-4}$ s$^{-1}$. Initial porosity: $f = 0.01$.

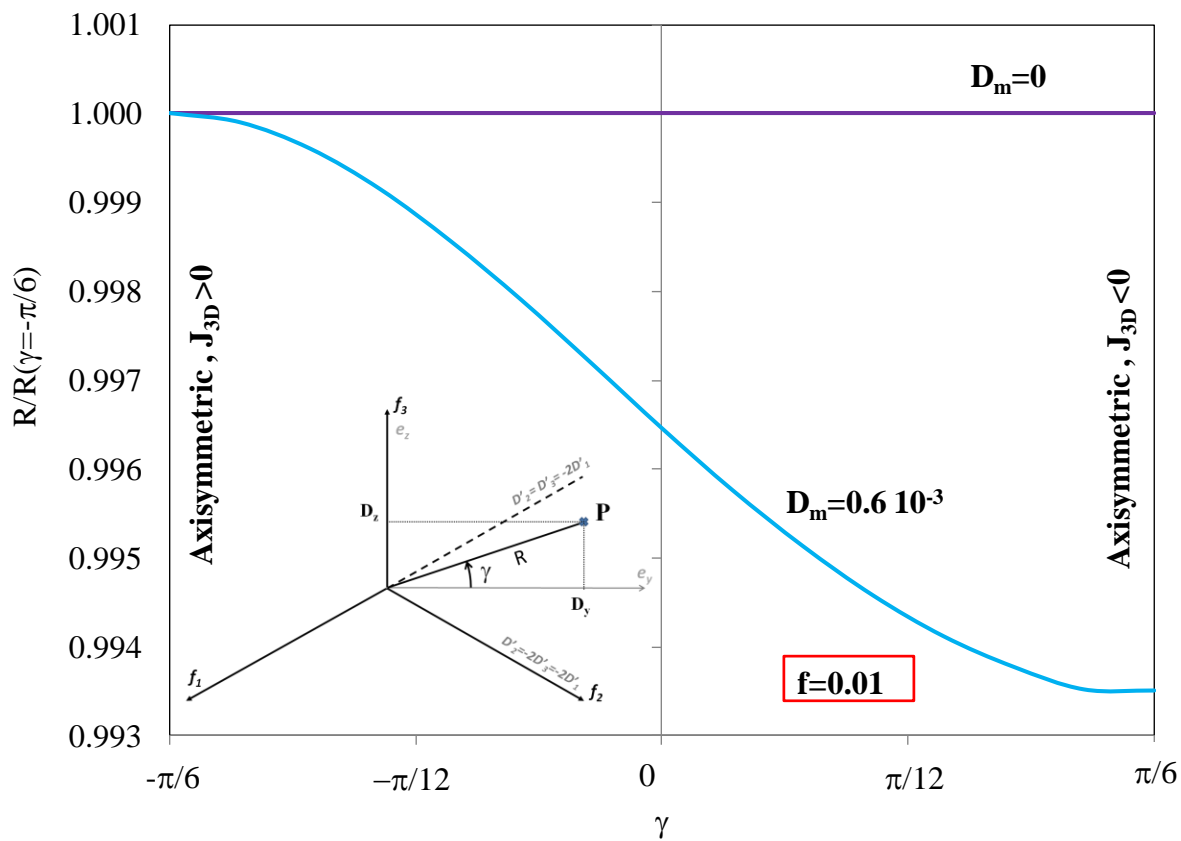

**Fig.5**. Evolution of $R(\gamma)$ (normalized by $R= R(-\pi/6)$) for the cross-section of the surface of the porous Mises material with the deviatoric planes $D_m = 6\cdot10^{-4}\,s^{-1}$ and $D_m = 0$ (von Mises behavior), respectively. Initial porosity: $f = 0.01$. Note that the response depends on the third-invariant as revealed by the very specific coupling between the invariants R ($=\sqrt{2J_{2D}}$) and $\gamma$ (measure of $J_{2D}$ and $J_{3D}$).

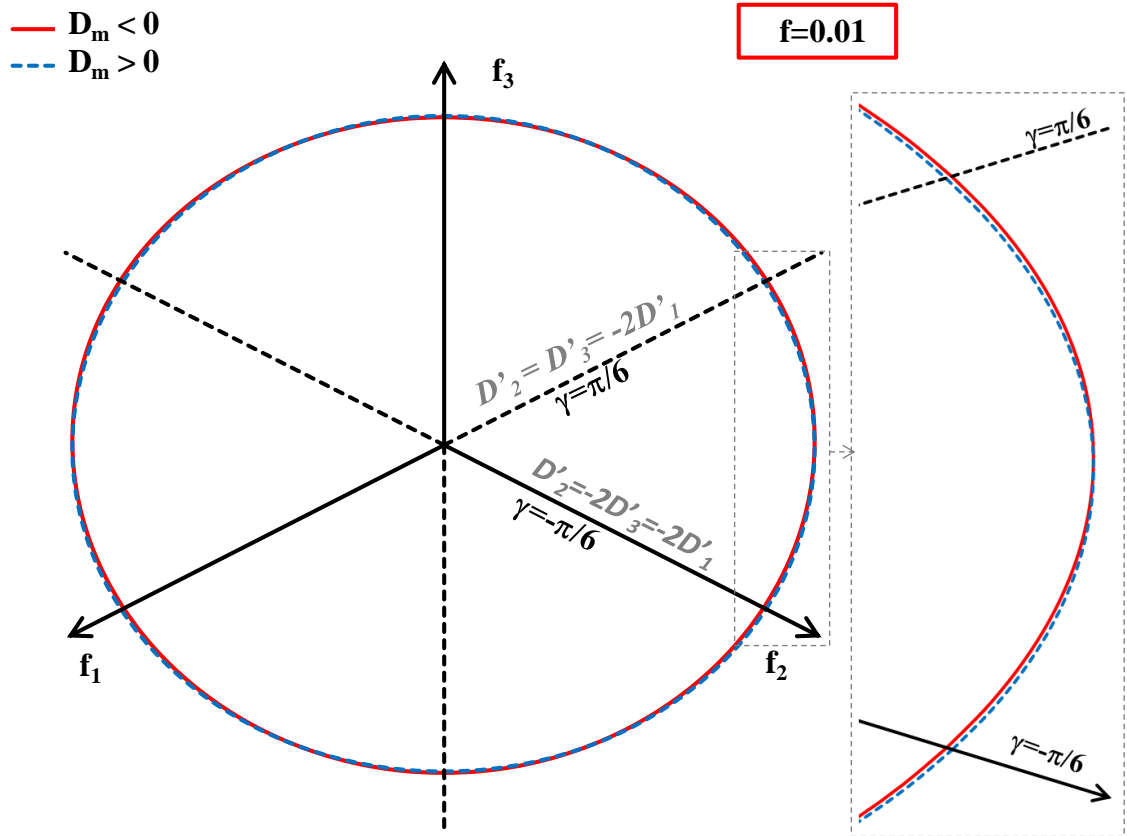

**Fig. 6**. Cross-sections of the surface of the porous von Mises material, $\Pi^{+}_{\text{Mises}}(\mathbf{D}, f) = 9.21 \cdot 10^{-3}$ (f =0.01) with the deviatoric planes $D_m = 6 \cdot 10^{-4}\,\text{s}^{-1}$ (interrupted line) and $D_m = -6 \cdot 10^{-4}\,\text{s}^{-1}$ (solid line). Note the centro-symmetry of the cross-sections due to the invariance of the plastic response to the transformation $(D_m, \mathbf{D}') \rightarrow (-D_m, -\mathbf{D}')$. Initial porosity: $f = 0.01$.

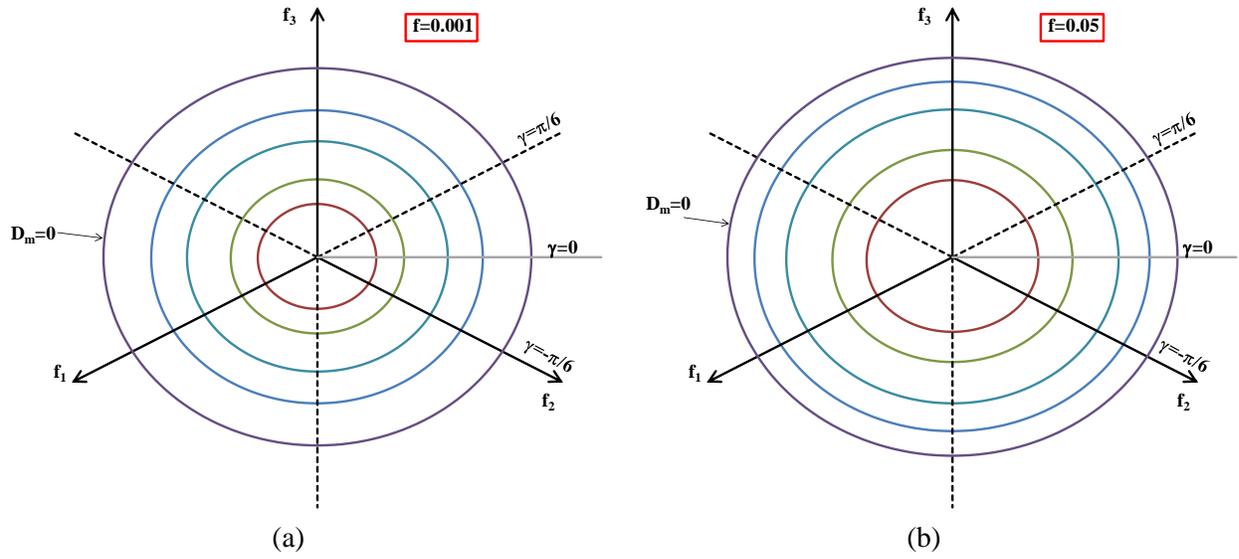

**Fig. 7**. Cross-sections of the 3-D isosurfaces of the porous Mises material with several deviatoric planes $D_m$ =constant for (a) porosity $f = 0.001$ and (b) porosity $f = 0.05$. In each case, the outermost cross-section corresponds to $D_m= 0$ while the innermost cross-section corresponds to $D_m/D_m^H = 0.9$, where $D_m^H$ represents the hydrostatic limit.

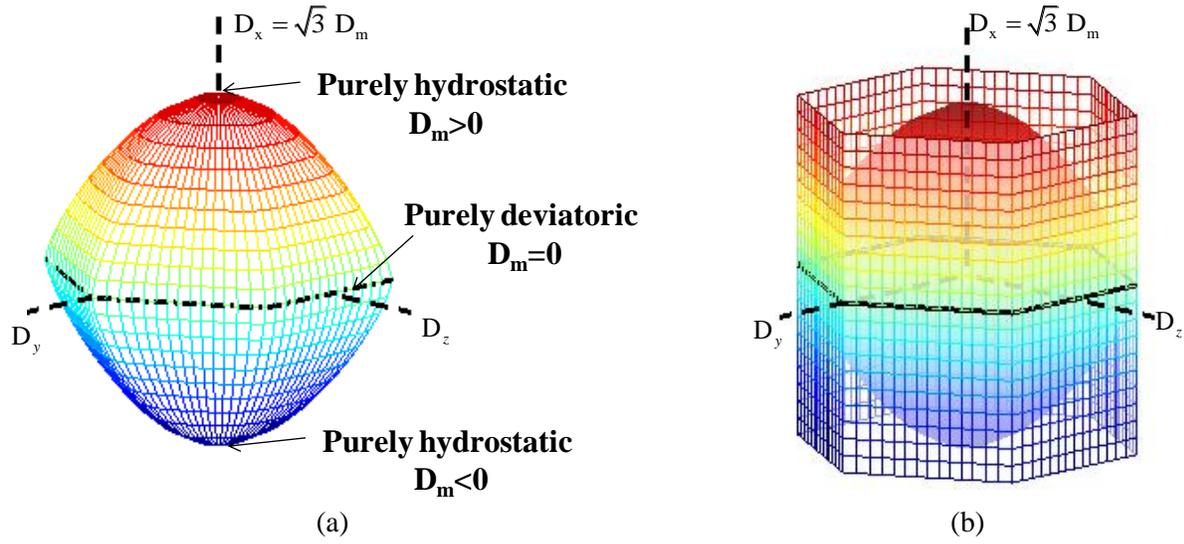

**Fig. 8**. The 3-D surface for a porous solid with Tresca matrix according to Eq. (24) for both tensile ($D_m = \text{tr}(\mathbf{D}) > 0$) and compressive ($D_m < 0$) states. Note that this convex surface contains all the points ($D_m$, R, γ) that produce the same plastic dissipation $\Pi^+_{\text{Tresca}}(\mathbf{D}, f) = 9.21 \cdot 10^{-3}$ for the porous solid. Initial porosity: $f = 0.01$. (b) Comparison between the 3-D isosurface of the porous material ($f = 0.01$) and that of the fully-dense material (Tresca behavior).

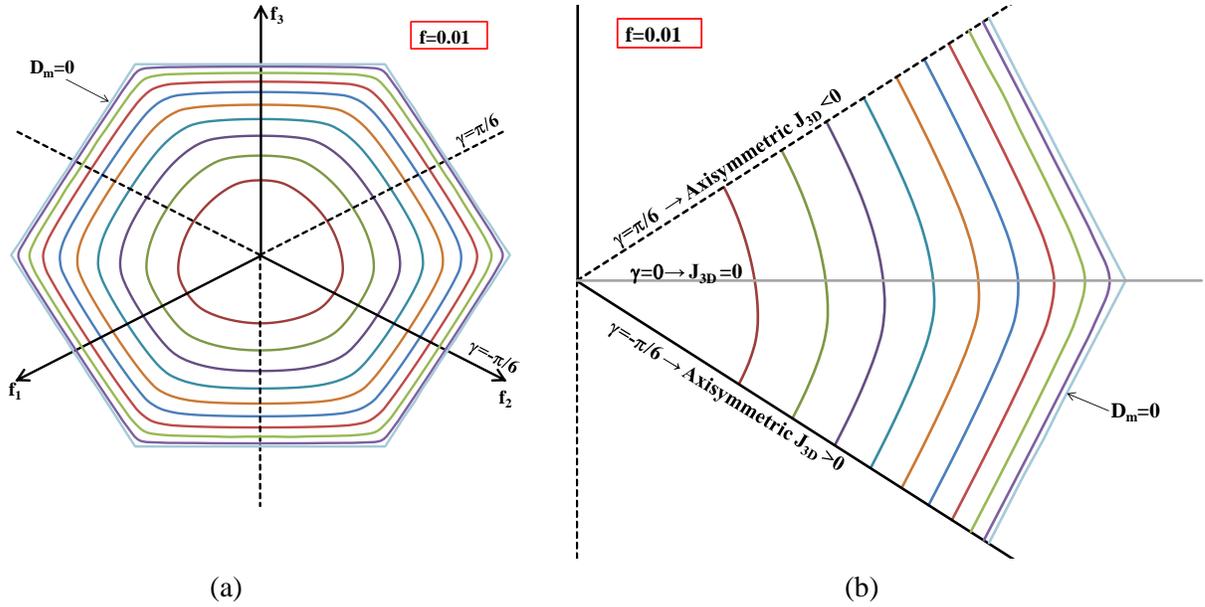

**Fig. 9**. Cross-sections of the 3-D isosurface of porous Tresca material (Eq. (25)) with several deviatoric planes $D_m$ =constant: Outer cross-section represents $D_m=0$ (matrix behavior) and the outmost inner cross-section corresponds to $D_m = 9 \cdot 10^{-4}$ s$^{-1}$: (a) Entire cross-section showing three-fold symmetry; (b) zoom in the sector: $-\pi/6 \leq \gamma \leq \pi/6$. Initial porosity: $f = 0.01$. Note the drastic change in the shape of the cross-section from a regular hexagon to smooth triangle, which indicates a very strong coupling between all three invariants.

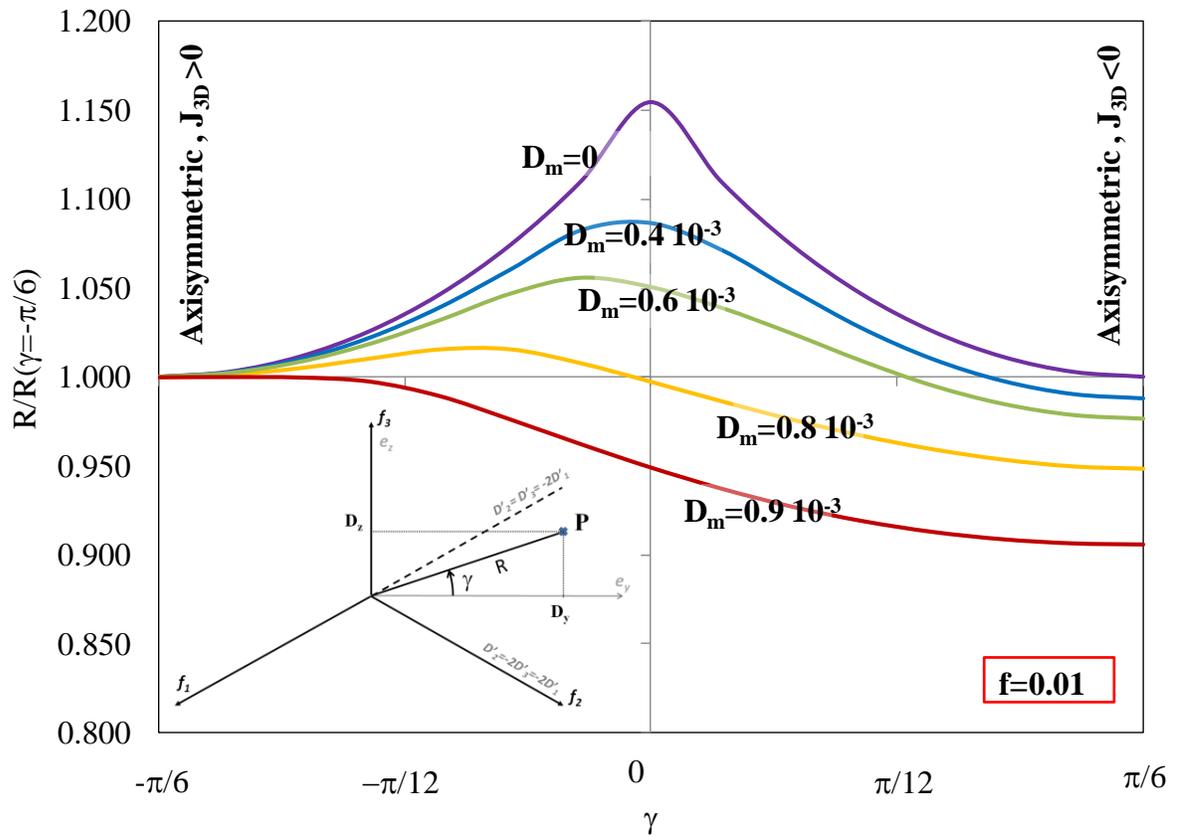

**Fig. 10**. Shape of the cross-sections revealed by the evolution of $R(\gamma)$ (normalized by $R(-\pi/6)$) corresponding to several deviatoric planes $D_m$ = constant for the porous Tresca material. Note that the plane $D_m$=0 represents the matrix behavior (i.e. Tresca). Initial porosity: $f = 0.01$. The very strong influence of $D_m$ on the couplings between the second and third-invariants (i.e. R= R($\gamma$)) is revealed.

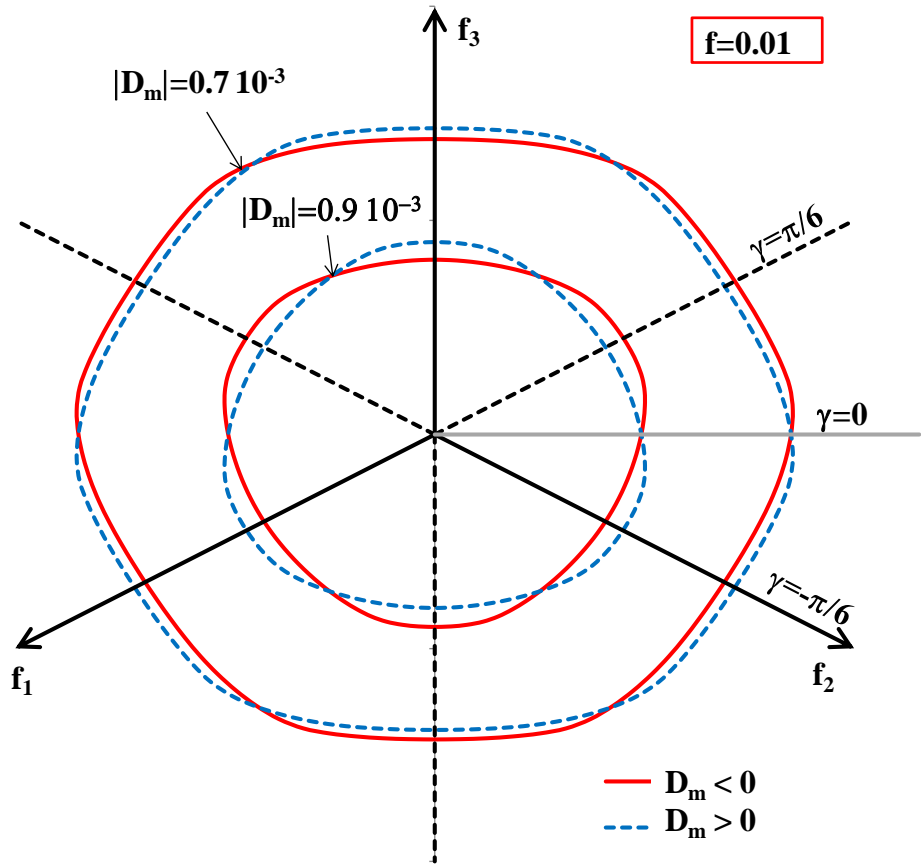

**Fig. 11**. Cross-sections of the surface of the porous Tresca material, $\Pi^+_{Tresca}(\mathbf{D}, f) = 9.21 \cdot 10^{-3}$ (f =0.01) with the deviatoric planes $D_m = 7 \cdot 10^{-4} \, s^{-1}$ and $D_m = 9 \cdot 10^{-4} \, s^{-1}$, respectively (interrupted lines) as well as the cross-sections with the planes $D_m = -7 \cdot 10^{-4} \, s^{-1}$ and $D_m = -9 \cdot 10^{-4} \, s^{-1}$, respectively (solid lines). Note the centro-symmetry of the cross-sections due to the invariance of the plastic response to the transformation $(D_m, \mathbf{D'}) \to (-D_m, -\mathbf{D'})$. Initial porosity: $f = 0.01$.

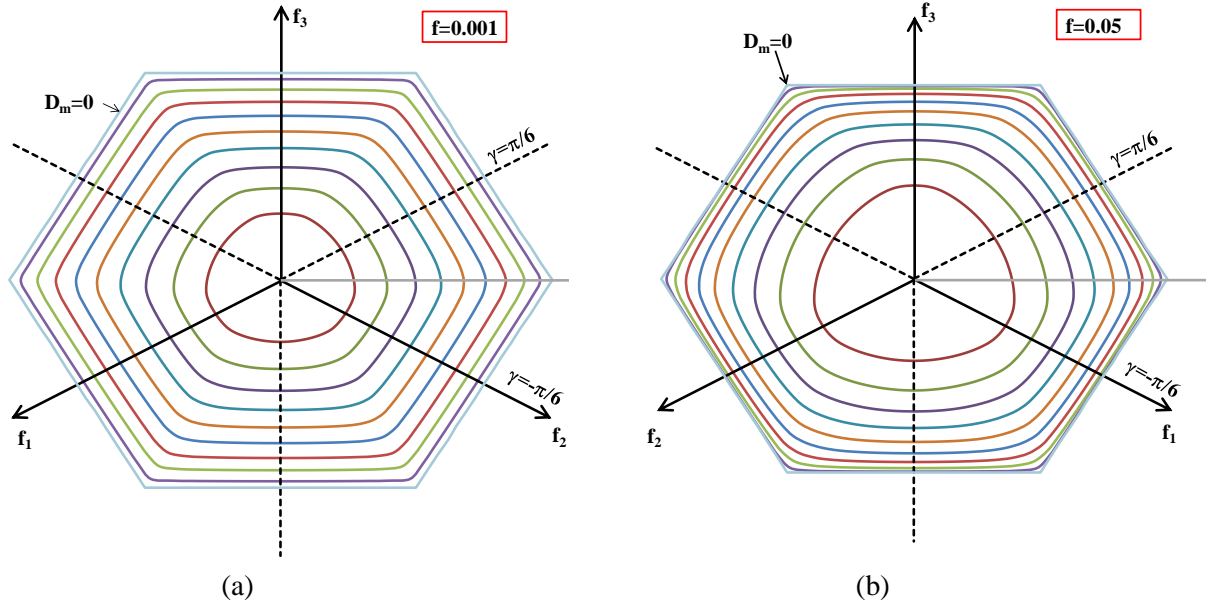

**Fig. 12**: Cross-sections of the 3-D isosurfaces of the porous Tresca material (Eq. (25)) with several deviatoric planes $D_m$ =constant for (a) porosity $f = 0.001$ and (b) porosity $f = 0.05$. In each case, the outermost cross-section corresponds to $D_m = 0$ while the innermost cross-section corresponds to $D_m/D_m^H = 0.9$, where $D_m^H$ represents the hydrostatic limit.

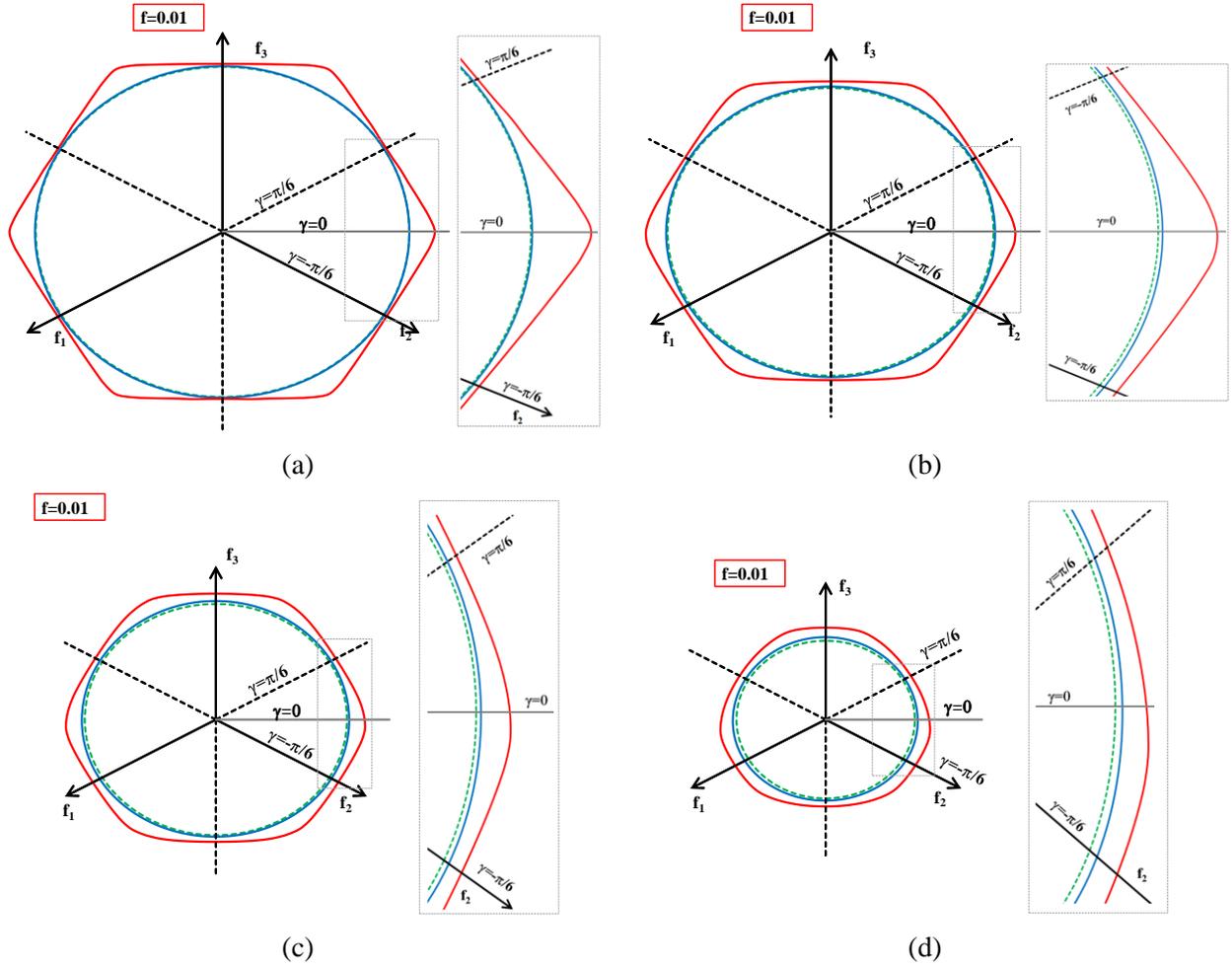

**Fig. 13**: Comparison between the shapes of the cross-sections of the SRP dual to Gurson (1977) Eq. (27), interrupted line), the exact SRP for a porous von Mises material (Eq.(19), blue solid line), and the SRP for a porous Tresca material (Eq.(25)) (red solid line) corresponding to the same porosity f = 0.01 and the same level of plastic energy 9.21 $10^{-3}$ . The cross-sections correspond to: (a) $D_m = 2 \cdot 10^{-4} s^{-1}$; (b) $D_m = 4 \cdot 10^{-4} s^{-1}$; (c) $D_m = 6 \cdot 10^{-4} s^{-1}$; (d) $D_m = 8 \cdot 10^{-4} s^{-1}$. Initial porosity: $f = 0.01$. Note that of the three potentials, Gurson's is the most dissipative ( interrupted line) while the porous Tresca SRP is the least dissipative (red solid line).

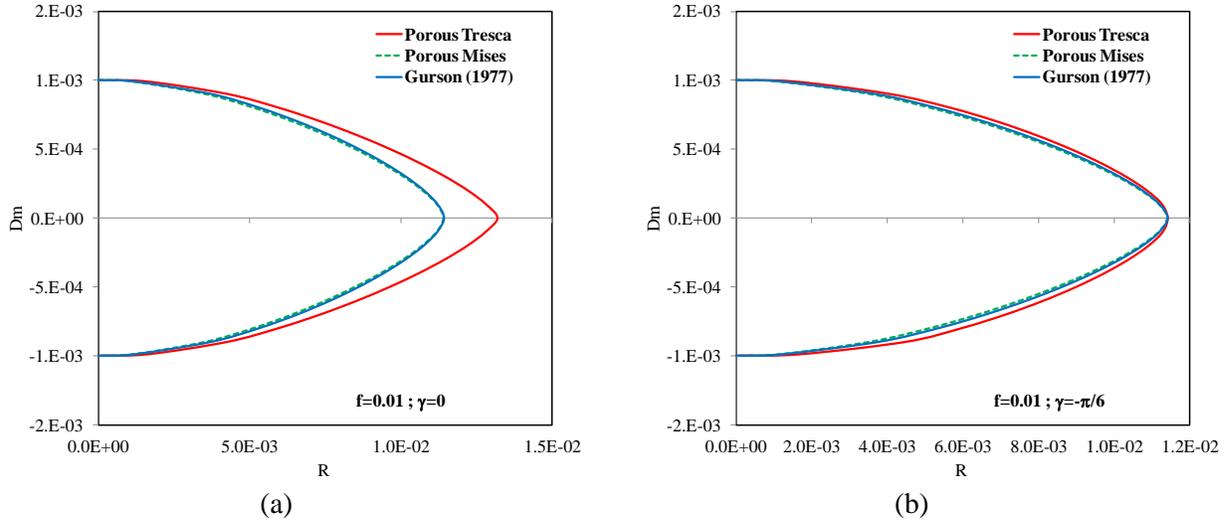

**Fig. 14**. Cross-sections of the 3-D isosurfaces of the porous Tresca material, porous Mises material, and Gurson's isosurface at $\gamma$ = constant: (a) $\gamma = 0$ ; (b) $\gamma = -\pi/6$. Initial porosity $f = 0.01$. Note that only for purely hydrostatic loadings (R=0), the response according with the three criteria is the same irrespective of $\gamma$.